\documentclass{optica-article}
\usepackage{caption}
\usepackage{subcaption}
\usepackage{comment}
\usepackage{soul}
\usepackage{graphicx}
\journal{oe}

\articletype{Research Article}

\usepackage{lineno}

\begin{document}

\title{Semi-analytical technique for the design of disordered  coatings with tailored optical properties}

\author{Bhrigu Rishi Mishra,\authormark{1} Nithin Jo Varghese,\authormark{1} and Karthik Sasihithlu\authormark{1,*}}

\address{\authormark{1}Department of Energy Science and Engineering, Indian Institute of Technology Bombay\\
}

\email{\authormark{*}ksasihithlu@iitb.ac.in} 



\begin{abstract}
Disordered media coatings are finding increasing use in applications such as day-time radiative cooling paints and solar thermal absorber plate coatings  which require tailored optical properties over a broad spectrum ranging from visible to far-IR wavelengths. Both monodisperse and polydisperse configurations with thickness of coatings up to 500 $\mu m$  are currently being explored for use in these applications. In such cases it becomes increasingly important to explore utility of analytical and semi-analytical methods for design of such coatings to help reduce the computational cost and time for design. While well-known analytical methods such as Kubelka-Munk and four-flux theory have previously been used for analysis of disordered coatings, analysis of their utility has so far in literature been restricted to either solar spectrum or IR but not simultaneously over the combined spectrum as required for the above applications. In this work, we have analysed the applicability of these two analytical methods for such coatings over the entire wavelength range from visible to IR, and based on observed deviation from exact numerical simulation we propose a semi-analytical technique to aid in the design of these coatings with significant computational cost savings.
\end{abstract}

\section{Introduction}
Disordered coatings, which consist of   dielectric/metal nanoparticles dispersed randomly in a matrix, find their application in several fields such as solar thermal absorber coatings \cite{Gunde2000}, solar reflecting coatings \cite{Nilsson1995}, color paints \cite{Quinten2001}, translucent paints \cite{Bandpay2018}, 
tissue optics \cite{Roy2012}, daytime passive radiative cooling coatings \cite{Huang2017,Bao2017,Mishra2021}, and many more.
The main advantages that such disordered media offer that make them an attractive proposition for use in these applications are their cost-effective means of fabrication, and tunability of the desired optical properties of the coating - since the spectral position of Mie (plasmon) resonance of the embedded dielectric (metal) particles strongly depend on the size of the particles.
The main challenging task in the design of such disordered media is the modelling of its optical  properties. 
Techniques based on homogenization of the composite structures that predict effective permittivity and permeability of the disordered media, such as Maxwell-Garnett theory \cite{Maxwell-Garnett1904} and Bruggeman's model \cite{Bruggeman1935}, are valid only when the particle sizes are much smaller than the incident wavelength \cite{Bohren1986}.
Doyle et al. \cite{Doyle1989} showed that the use of Mie coefficients in this effective medium theory provides good accuracy to the calculation of effective optical properties of metal spheres suspended in a polymer. However, the theory predicts absorption for a nonabsorbing particle \cite{Bohren1986} and thus  cannot be used to predict the effective refractive index of disordered media for solar reflecting paint/coatings where non absorbing particles are utilized. In literature, other analytical techniques developed for this objective include those which consider diffusion of photons \cite{Gate1971}, and those which solve the radiative transfer equation under $N$-flux ($ 2\leq N \leq 4$) approximations \cite{ishimaru1978wave,caron2004radiative}. Of these methods Kubelka Munk (KM) theory \cite{KM1931} (for which $N=2$) and the four-flux (FF) method \cite{Maheu1984} are commonly used. In addition, simulation techniques such as the Monte Carlo method \cite{MC}, and exact electromagnetic solvers are employed to model the optical/radiative properties of disordered coatings.
However, these simulation techniques do not present a clear picture linking the microscopic properties of the particles, such as the scattering and absorption coefficients,  to the macroscopic optical properties  of the coating. Moreover, exact electromagnetic solvers 
which solve Maxwell's equations numerically to obtain radiative properties of the coating,
put a premium on computational resources and the  time for design when the thickness of random media is in the order of tens/hundreds of microns - as is currently being deployed in these applications. Particularly when several parameters of the configuration are in play, such as that encountered in disordered media, analytical techniques such as KM and FF theories  provide important means to arrive at an optimum combination of the parameters with minimal computational resources while also explicitly linking the properties of the micro constituents to the observed optical properties of the coating. 

KM and FF theories have so far in literature been used in applications where the spectrum of interest has been limited to either the visible spectrum or IR separately. 
For example, 
KM theory has been used extensively in paints and coatings \cite{Gunde2000,Quinten2001}, paper industry \cite{Dzimbeg2011}, tissue optics \cite{Roy2012} among others. Similarly, the FF method has been used extensively by researchers to model, predict and optimize the optical properties of light scattering coatings \cite{ Nilsson1995, Etherden2004, Genty-Vincent2017, Gali2017a}. 
However, the applicability of these theories  over a broad spectrum covering both visible and IR spectrum simultaneously has not been a subject of attention. This becomes important when designing coatings for applications such as day-time passive radiative cooling and solar thermal absorber plate coatings where tailored optical properties over a broad spectrum covering both the solar spectrum as well as far-infrared are crucial. For example, coatings for day-time passive radiative cooling \cite{Raman2014}  require  high reflectivity in solar spectrum (0.3-3~$\mu$m wavelength range) and high emissivity in  the infra-red spectrum (5-15~$\mu$m wavelength range). It is not obvious that the analytical techniques  retain their accuracy over such a broad spectrum since with increasing wavelength there is a possibility that the nature of scattering transitions from the independent scattering regime (where scattering cross-section of particles can be superposed) to dependent scattering regime (where near-field effects, and interference between far-field scattered radiation become important). Previously reported relation \cite{Hottel1971} demarcating the two regimes has been obtained  from experimental observations carried out in the visible spectrum only. Thus there is a need to explore the applicability of these analytical techniques over a broad spectrum in greater depth. In regimes where the predictions from these analytical techniques are not satisfactory, other possible methods of design which combine the accuracy of exact electromagnetic solvers with the minimal computational requirements of  analytical methods are expected to be of pressing need to researchers interested in designing such coatings. 

With this in mind, the manuscript has been arranged as follows. In Section \ref{sec:analytical} we have compared the reflectivity and emissivity predictions of KM and FF techniques with results from exact numerical solvers for different degrees of absorption in the particles  (imaginary index of particle = 0.0, $10^{-2}$, and $10^{-1}$) and in the matrix (imaginary index of matrix = 0.0, $10^{-4}$, and $10^{-2}$) for different thickness of the coating (10~$\mu$m and 50~$\mu$m) in the wavelength range $0.3 - 15~\mu m$.    We show that these techniques are accurate over the entire spectrum when particles are in the limit of independent scattering and under low absorption conditions but fail when volume fraction of particles is high such that interaction among particles is no longer negligible or when absorption in the matrix/particles is high. For such conditions where analytical techniques fail to predict the optical properties accurately we propose an alternative technique which combines the use of exact numerical solver and KM theory which we show can predict optical properties with accuracy as well as with minimal computational requirements. This `semi-analytical' technique is detailed in Section \ref{sec:semi_anyl}. In the end, as an example to showcase the applicability of this semi-analytical technique,  we predict the properties of a disordered coating  suitable for the application of passive radiative cooling 
and compare these with experimental  measurements  previously reported in literature.

\section{Analytical techniques - Kubelka Munk (KM) and the Four-Flux (FF) methods}
\label{sec:analytical}
\begin{figure}[h!]
     \centering
     \includegraphics[width=7cm]{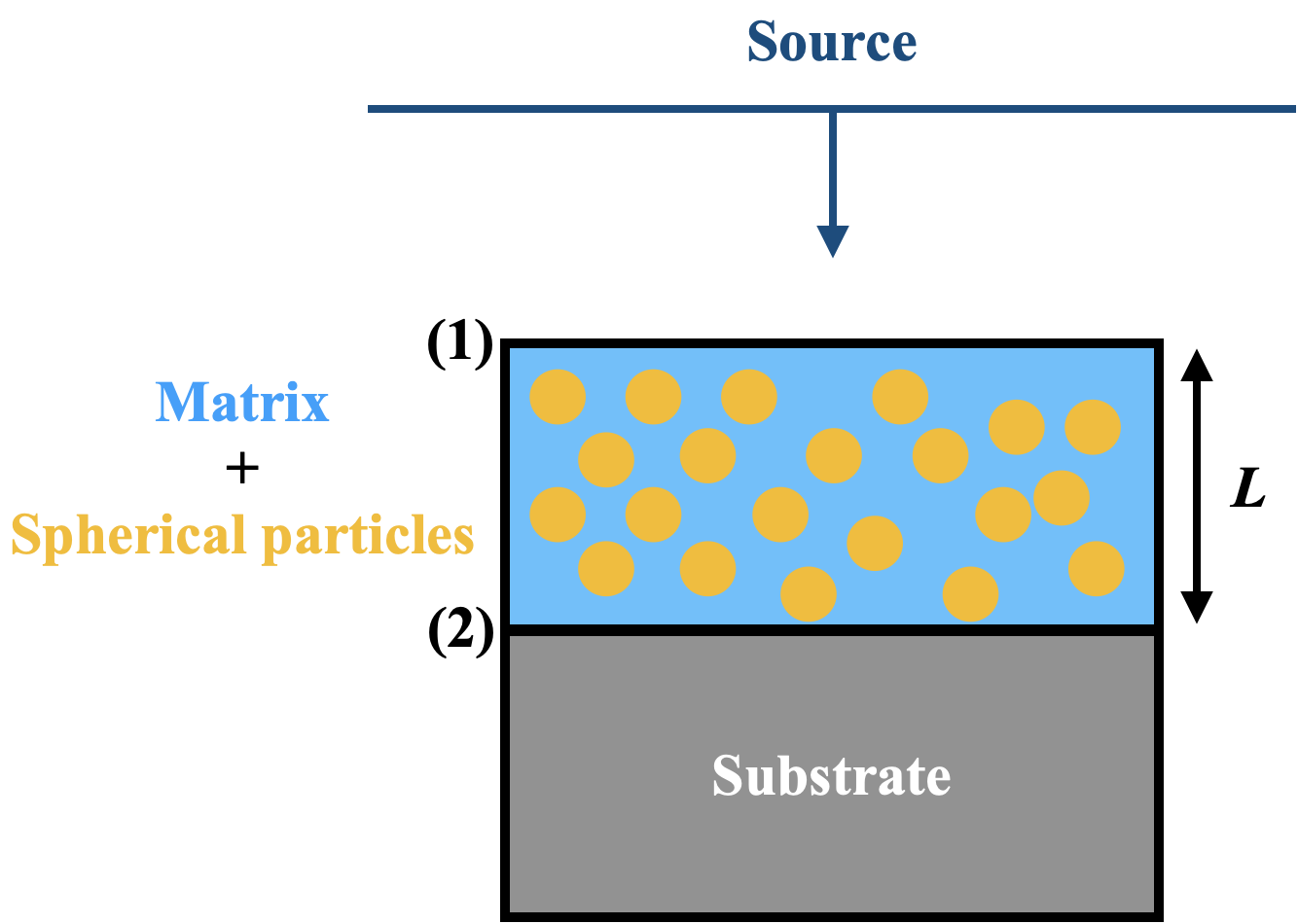}
    \caption{Schematic of coating considered in this work with incident plane wave source.
    }
    \label{fig:schematic_paper}
\end{figure}

We start with the expressions for reflectivity and transmissivity of the coating as obtained from KM and FF theories  which we use in the work to analyze the optical properties of the disordered coating. Detailed derivations  of these expressions can be found in several references \cite{Kubelka1947,Maheu1984,Maheu1986}. The optical coating considered in this work is a plane-parallel slab of  particulate composite on a substrate as shown in Fig. \ref{fig:schematic_paper}. The composite is considered to be of finite thickness and infinite extension in the lateral direction. The randomly distributed spherical particles  embedded within the host medium (also called the matrix) act as inhomogeneities to the propagating EM wave, thereby causing its scattering (and absorption, in case the particle is lossy). The objective is to predict the optical properties of this coating including the total reflectance, transmittance and absorption.
\\
The expression for the reflectivity ($R_{\text{KM}}$) and transmissivity ($T_{\text{KM}}$) from KM theory is  given by \cite{Kubelka1947,Quinten2001}:
\begin{equation}
    R_{\text{KM}}=\frac{(1-\gamma)(1+\gamma)(\exp(AL)-\exp(-AL))}{(1+\gamma)^2\exp(AL)-(1-\gamma)^2\exp(-AL)}
    \label{eq:KM_R}
\end{equation}
\begin{equation}
  T_{\text{KM}}=\frac{4\gamma}{(1+\gamma)^2\exp(AL)-(1-\gamma)^2\exp(-AL)}
  \label{eq:KM_T}
\end{equation}
where, $L$ is the thickness of the layer,  the coefficients $\gamma$ and $A$  are given by \cite{Quinten2001, Molenaar1999}:
\begin{equation}
\label{gamma,A}
\gamma = \sqrt{k/(  k + s')} ;\,\, A = \sqrt{2k(2k+s')} 
\end{equation} 
with 
\begin{equation}
    s'=\frac{3s(1-g)-k}{4}; \,\, 
    \label{eq:SK}
\end{equation}
and the factors $s$ and $k$ got using Mie theory \cite{Gunde2000}
\begin{equation}
    s=\frac{3fQ_{\text{sca}}}{4r}; \,\, \,\,\,\, k=\frac{3fQ_{\text{abs}}}{4r},
    \label{eq:Qsca_Qabs}
\end{equation}
where $f$ is the volume fraction, $r$ is the radius of the sphere, $Q_{\text{sca}}$ $(Q_{\text{abs}})$ is the  Mie scattering (absorption) efficiency of a single particle embedded in a host medium of index $n_h$, and $g$ is the asymmetry parameter. Expressions for $Q_{\text{sca}}$, $Q_{\text{abs}}$, and $g$ in terms of standard Mie coefficients can be found in Ref. \cite{bohren2008absorption}. It should be pointed out that the relations  between the coating properties $\gamma$ and $A $, and the particle properties $s$ and $k$ given in Eq. \ref{gamma,A} are not unique -  several other relations \cite{Vargas1997,Vargas2002,Yang2004,Murphy2006,Bandpay2018, Roy2012,Thennadil2008} have been proposed over the years. 
The expressions in Eq. \ref{gamma,A} and Eq. \ref{eq:SK}, taken from Ref. \cite{Quinten2001, Molenaar1999}, is representative and have been chosen for demonstrative reasons. As we will see in Sec. \ref{sec:semi_anyl} the semi-analytical method being proposed in this work does not depend on such relations and hence do not affect the central results of this work. %

In the limit of low absorption, $kL\rightarrow 0$, the reflectivity in Eq. (\ref{eq:KM_R}) can be shown  to reduce to \cite{Kubelka1947}:
\begin{equation}
    R_{\text{KM}} = \dfrac{s'L}{s'L + 1}.
    \label{eq:KM2}
\end{equation}
%
It must be noted that Eq. (\ref{eq:KM_R}) and Eq. (\ref{eq:KM_T}) do not take into account surface reflection of incident radiation at interface (1). Modified reflectance $R_0$ and transmittance $T_0$ which take into account surface reflection correction are calculated using \cite{Saunderson1942}:
\begin{equation}
    R_\text{0} = R_\text{c} + \frac{(1-R_\text{c})(1-R_\text{i})R_{\text{KM}}}{1-R_\text{i} R_{\text{KM}}}; \,\,\,\,\, T_\text{0} = \frac{(1-R_\text{c})T_{\text{KM}}}{1-R_\text{i} R_{\text{KM}}}
    \label{eq:KM_Rsurf}
\end{equation}
where, $R_\text{c}$ is the specular reflectance of  incident light got from Fresnel reflection which for normal incidence from a medium of index $n_{\text{surr}}$  
reads:
\begin{equation}
R_\text{c} = \left( \frac{n - 1}{n + 1} \right)^2
\end{equation}
with $n = n_\text{h}/n_{\text{surr}}$ and $R_\text{i}$ is the diffuse reflectance of internal radiation at  interface (1), marked in Fig. \ref{fig:schematic_paper}, which is calculated using:
\begin{equation}
    R_i = 2\int_{0}^{\pi/2} \rho(\theta)\,\text{sin}\theta \, \text{cos}\theta \, \text{d}\theta.
    \label{eq:diffuse_ref}
\end{equation}
where,
 from Fresnel's coefficients:
\begin{equation}
    \rho (\theta) = \frac{1}{2}\left[\left(\dfrac{\sqrt{n^2-\text{sin}\theta}-\text{cos}\theta}{\sqrt{n^2-\text{sin}\theta}+\text{cos}\theta}\right)^2+\left(\dfrac{n^2 \text{cos}\theta-\sqrt{n^2-\text{sin}\theta}}{n^2 \text{cos}\theta+\sqrt{n^2-\text{sin}\theta}}\right)^2\right].
    \label{eq:Fresnel}
\end{equation}
%
The expression for $R_\text{i}$ from Eq. \ref{eq:diffuse_ref} can be used even the limit of low diffuse scattering since the contribution from the product $R_{i} R_{\text{KM}}$ will be negligible in this regime. %

 Many configurations developed for radiative cooling application \cite{Bao2017,atiganyanun2021use,Zhang2021Zro2} and solar absorber plates \cite{cao2014review,wang2018design} involve use of a substrate. 
 In the presence of a substrate, the net reflectance and transmittance from Eq. \ref{eq:KM_Rsurf} will have to be further modified as \cite{Kortum1969}:
\begin{equation}
    R = R_\text{0} + \frac{T_\text{0}^2 R_\text{g}}{1-R_\text{0} R_\text{g}}; \,\,\,   T = \frac{(1-R_\text{g})T_\text{0}}{1- R_\text{0} R_\text{g}}
    \label{eq:KM_Rb}
\end{equation}
Here $R_\text{g}$ is the diffuse reflectance at interface (2) obtained from Eq. \ref{eq:diffuse_ref} with $n = n_\text{h}/n_\text{g}$. 
The substrate index $n_g$ is taken to be 1.5 in this work. 

%
%
The derivation of reflection and transmission coefficients from KM theory assumes that the incident light is diffuse. When the incident radiation is collimated, alternate methods such as the four-flux theory, which take into account the propagation of both collimated and diffuse radiation across the interfaces in two directions, are expected to be more accurate. This careful consideration of both collimated and diffuse components leads to expressions for the optical properties being far more complicated than in KM theory. The net reflection and transmission coefficients when incident radiation is fully collimated can be expressed in terms of a summation over 
collimated-collimated reflectivity $(R_{\text{cc}})$, collimated-diffuse reflectivity $(R_{\text{cd}})$,  collimated-collimated transmissivity $(T_{\text{cc}})$, and collimated-diffuse transmissivity $(T_{\text{cd}})$ as:
\begin{equation}
    R = R_{\text{cc}}+R_{\text{cd}}+R_{\text{dd}}; \,\,\,\,  T = T_{\text{cc}}+T_{\text{cd}}+T_{\text{dd}}
    \label{eq:FF_R}
\end{equation}
Expressions for $R_{\text{cc}}$, $R_{\text{cd}}$, $T_{\text{cc}}$ and $T_{\text{cd}}$ are quite elaborate and  have been included in the supplementary document (Section S1) for reference. 

\begin{figure}[ht]
     \centering
     \begin{subfigure}[b]{0.49\textwidth}
    \centering
               \includegraphics[width=7cm]{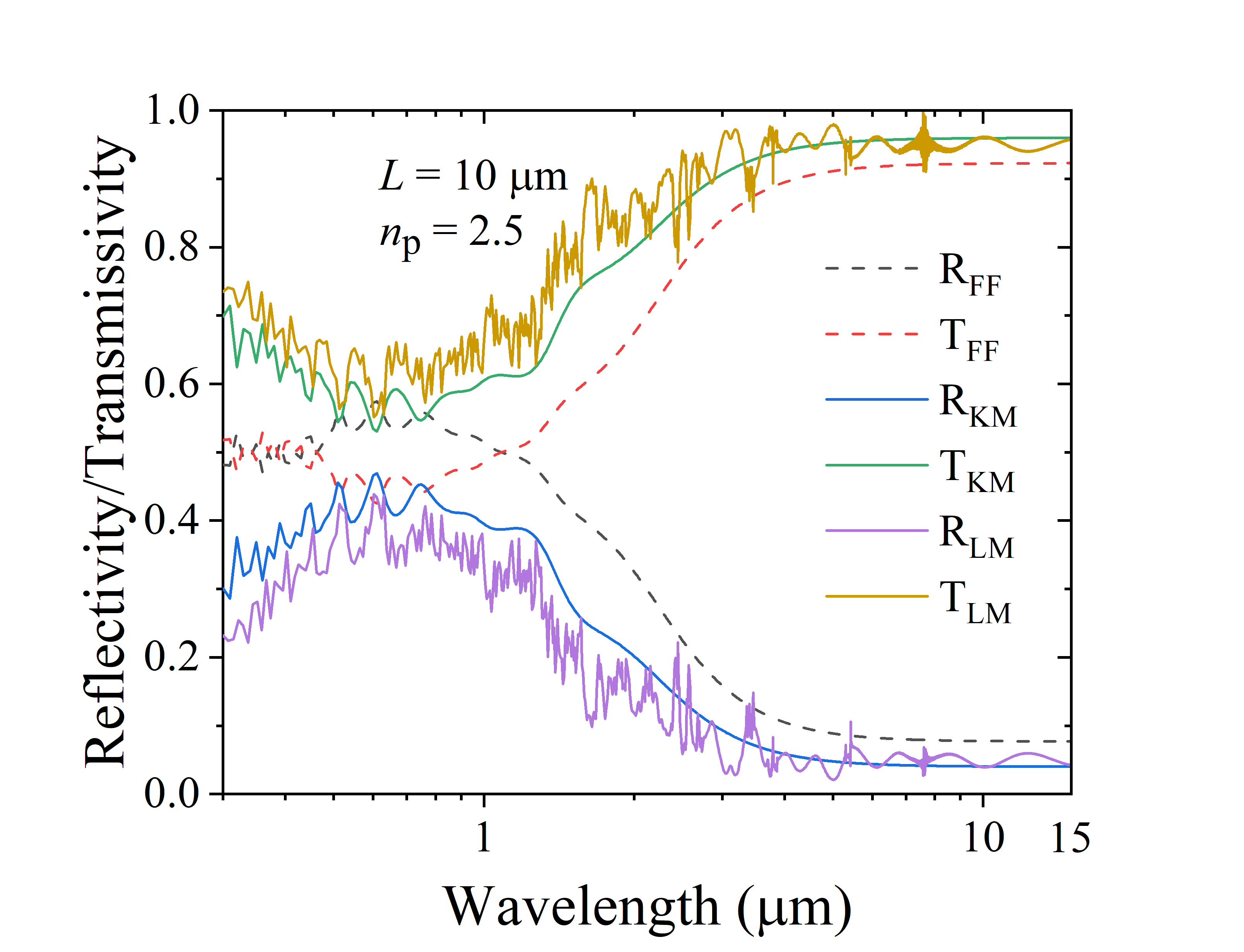} 
    \caption{}
    \label{fig:NAB_sr1}
     \end{subfigure}
     \hfill
     \begin{subfigure}[b]{0.49\textwidth}
    \centering
        \includegraphics[width=7cm]{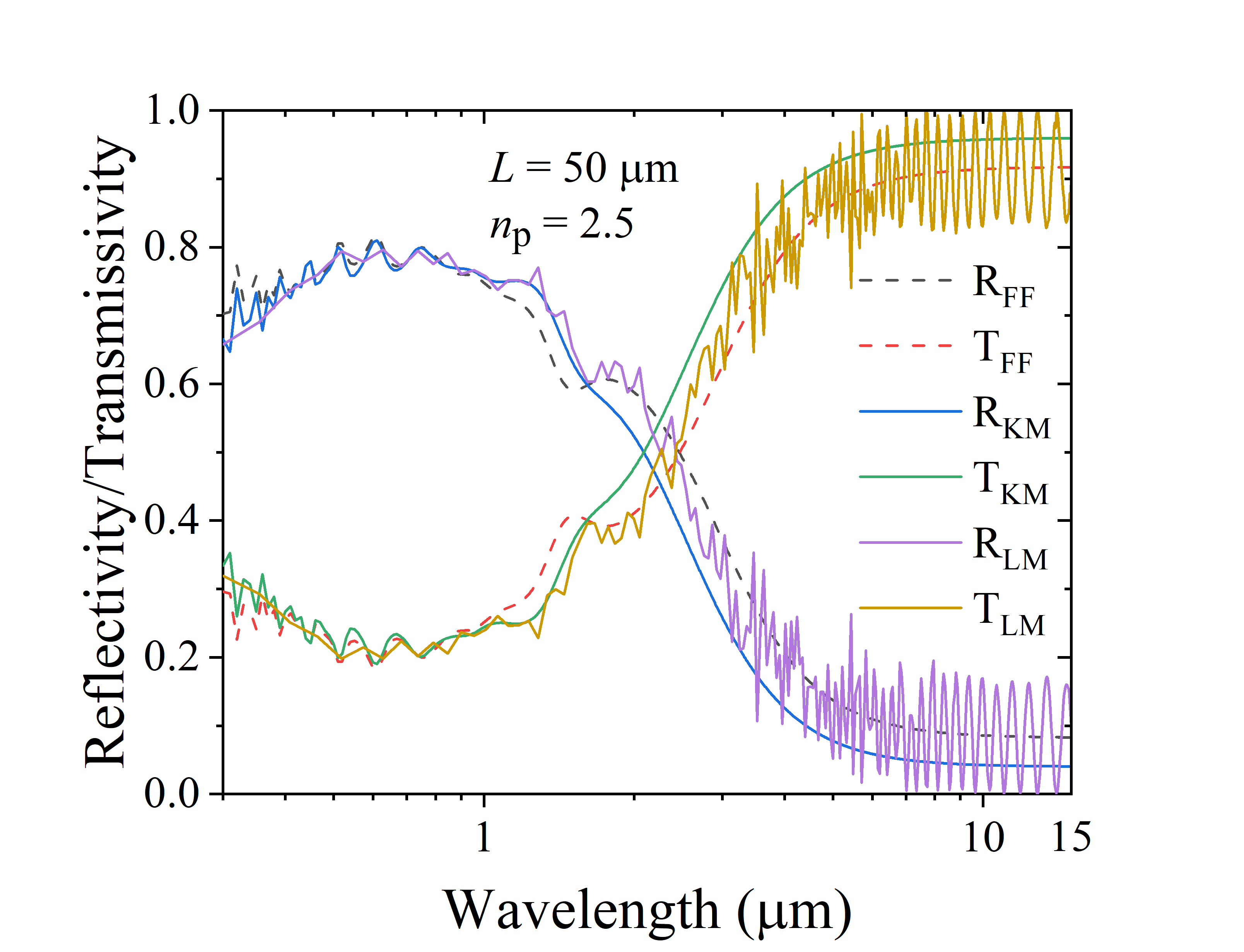}
    \caption{}
    \label{fig:NAB_sr2}
     \end{subfigure}
     \hfill
     \begin{subfigure}[b]{0.49\textwidth}
    \centering
        \includegraphics[width=7cm]{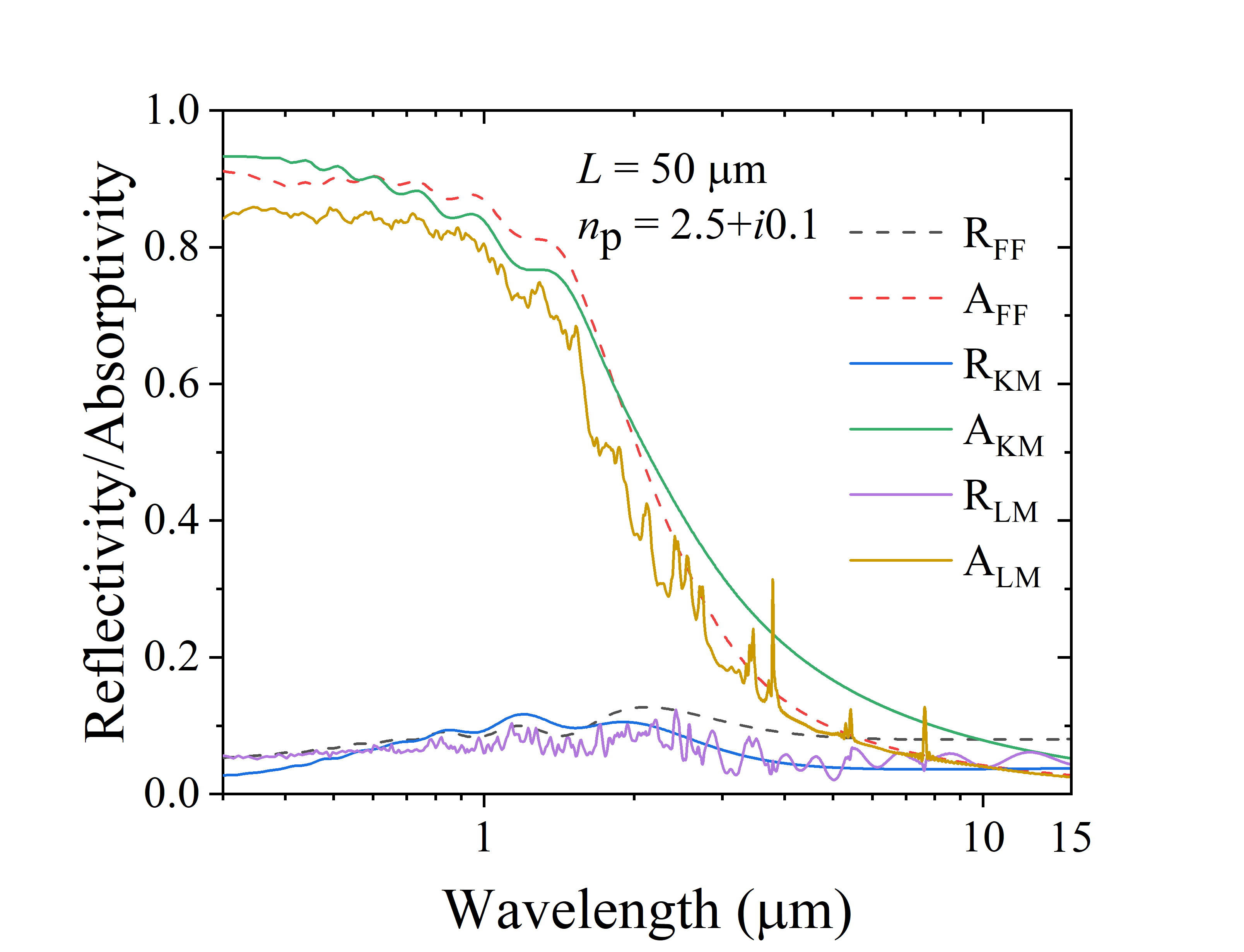}
    \caption{}
    \label{fig:NAB_sr5}
     \end{subfigure}
     \hfill
     \begin{subfigure}[b]{0.49\textwidth}
    \centering
        \includegraphics[width=7cm]{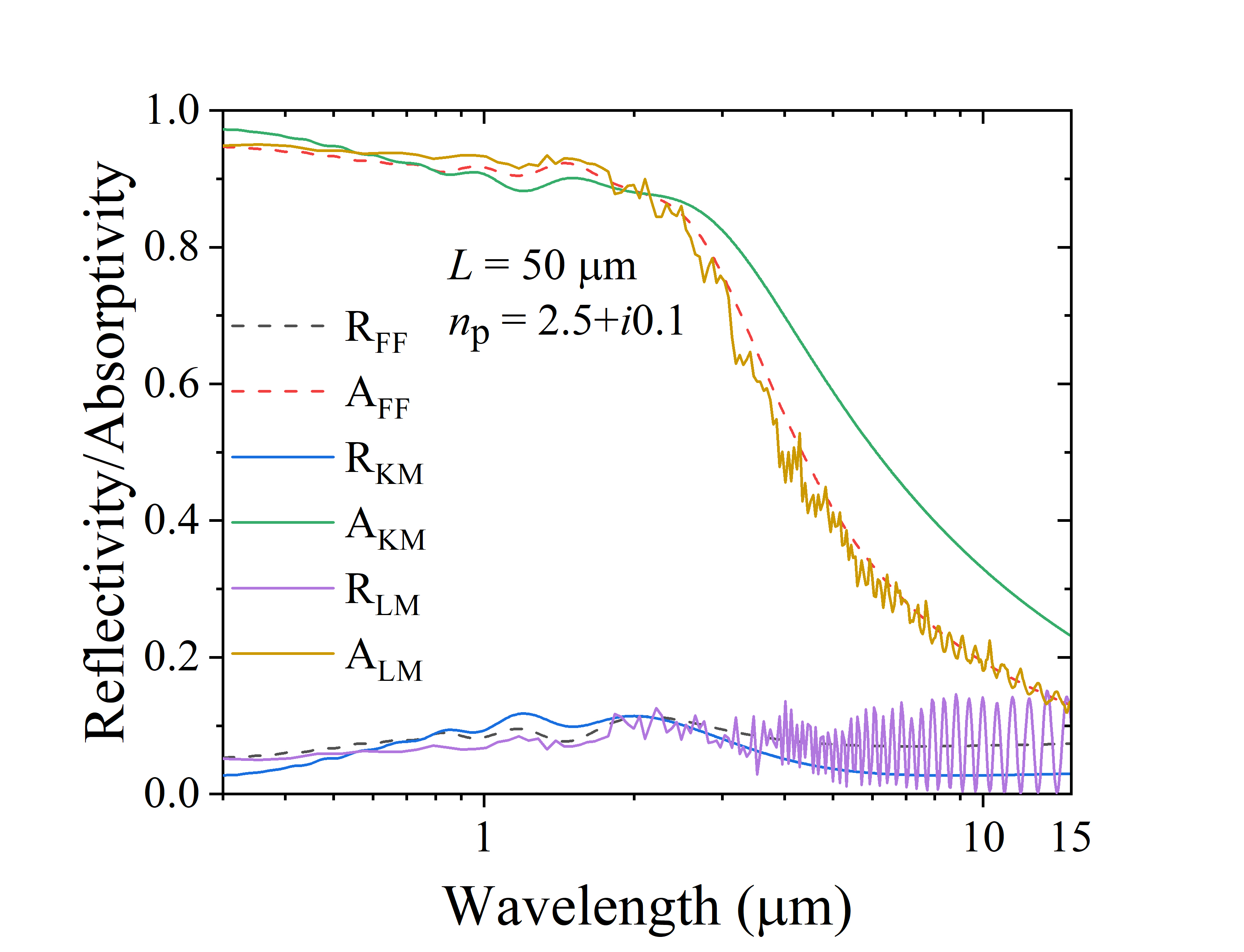}
    \caption{}
    \label{fig:NAB_sr6}
     \end{subfigure} 
    \caption{Reflectivity and transmissivity spectrum for $n_{\text{h}}=1.5,$ $r=0.25~\mu\text{m},$ $f=0.05,$ and (a) $n_{\text{p}}=2.5,$ $L=10~\mu\text{m}$; (b) $n_{\text{p}}=2.5,$ $L=50~\mu\text{m}$. Reflectivity and absorptivity spectrum for $n_{\text{h}}=1.5,$ $r=0.25~\mu\text{m},$ $f=0.05,$ and (c) $n_{\text{p}}=2.5+i0.1,$ $L=10~\mu\text{m}$; (d) $n_{\text{p}}=2.5+i0.1,$ $L=50~\mu\text{m}$.  Here, FF stands for four flux, KM for Kubelka Munk, and LM for Lumerical.}
    \label{fig:NAB_sr1_2}
\end{figure}

In Sections \ref{sec:mono}, \ref{sec:poly}, and \ref{sec:dep_scat}, we use the expressions for $R$ and $T$ given in Eqs. \ref{eq:KM_Rb} for KM theory and Eqs. \ref{eq:FF_R} for FF to predict the optical properties of disordered coatings and compare these with the results obtained from Lumerical FDTD solver \cite{Lumerical}. We analyze situations where both the particles and host medium are absorbing as well as non-absorbing, and also consider the effect of different thickness of the coating. The degree of absorption in particles considered in this work are relevant for dielectric inclusions typically included in coatings for use in radiative cooling and solar thermal applications. In addition,  to facilitate the parametric study we assume non-dispersive form of refractive index for both the particles as well as the host matrix. We  first confine our analysis to the independent scattering regime in Sections \ref{sec:mono} and \ref{sec:poly}, and  extend the analysis to dependent scattering regime later in Sec. \ref{sec:dep_scat}.
The FDTD simulations were set up in ANSYS Lumerical.  Periodic boundary conditions were applied in the lateral  $x$ and $y$ directions, and coating is illuminated with a plane wave source from $z$ direction. 
A mesh size of 30~nm was used which we find is sufficient for convergence (mesh convergence study is shown in supplementary Fig. S3). 

\subsection{Comparison of predictions from  KM, FF theories and FDTD solver in the independent scattering regime for monodisperse inclusions with and without absorption in particles and in host medium}
\label{sec:mono}

Figure \ref{fig:NAB_sr1} and \ref{fig:NAB_sr2} show the  comparison between KM, FF, and FDTD results for the case when particles are non-absorbing  and Fig. \ref{fig:NAB_sr5} and \ref{fig:NAB_sr6} show the corresponding comparison when particles are absorbing with imaginary index of particles $n_p'' = 10^{-1}$. We compare the predictions for different coating thicknesses $10~\mu m$ and $50~ \mu m$ keeping the other parameters $r = 0.25~\mu m$, $f=0.05$, $n_h = 1.5$ non varying. 
It is observed that particularly for smaller thickness of coating and in the absence of absorption the predictions from FF method deviates significantly from the FDTD simulations as compared to KM method in both the visible as well as IR spectrum. However, for larger thicknesses of the coating and in the presence of absorption in particles, FF is relatively more accurate than the KM method across the spectrum, more so for higher wavelengths. 

\begin{figure}[ht]
     \centering
     \begin{subfigure}[b]{0.49\textwidth}
    \centering
       \includegraphics[width=7cm]{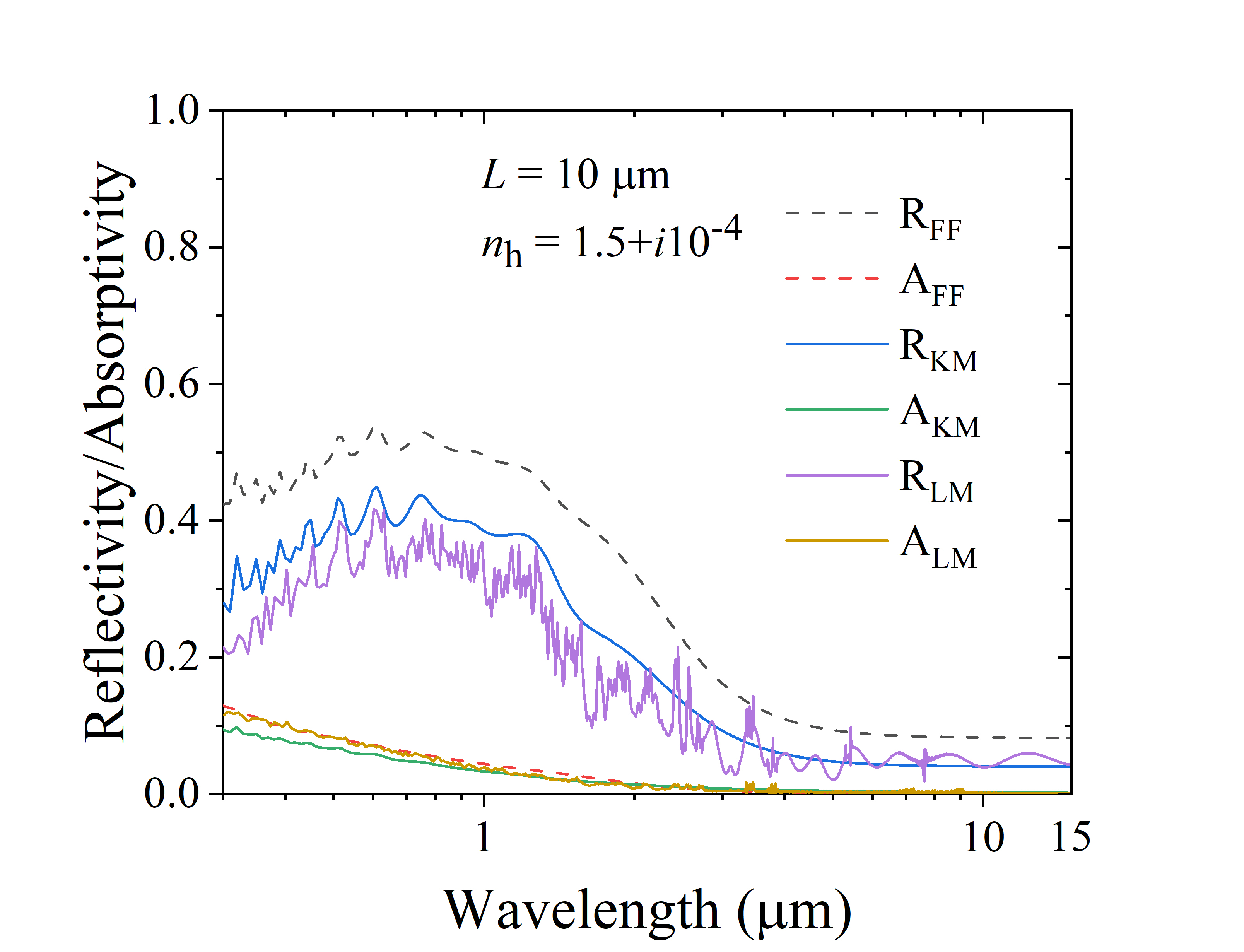} 
    \caption{}
    \label{fig:AB_sr1}
     \end{subfigure}
     \hfill
     \begin{subfigure}[b]{0.49\textwidth}
    \centering
        \includegraphics[width=7cm]{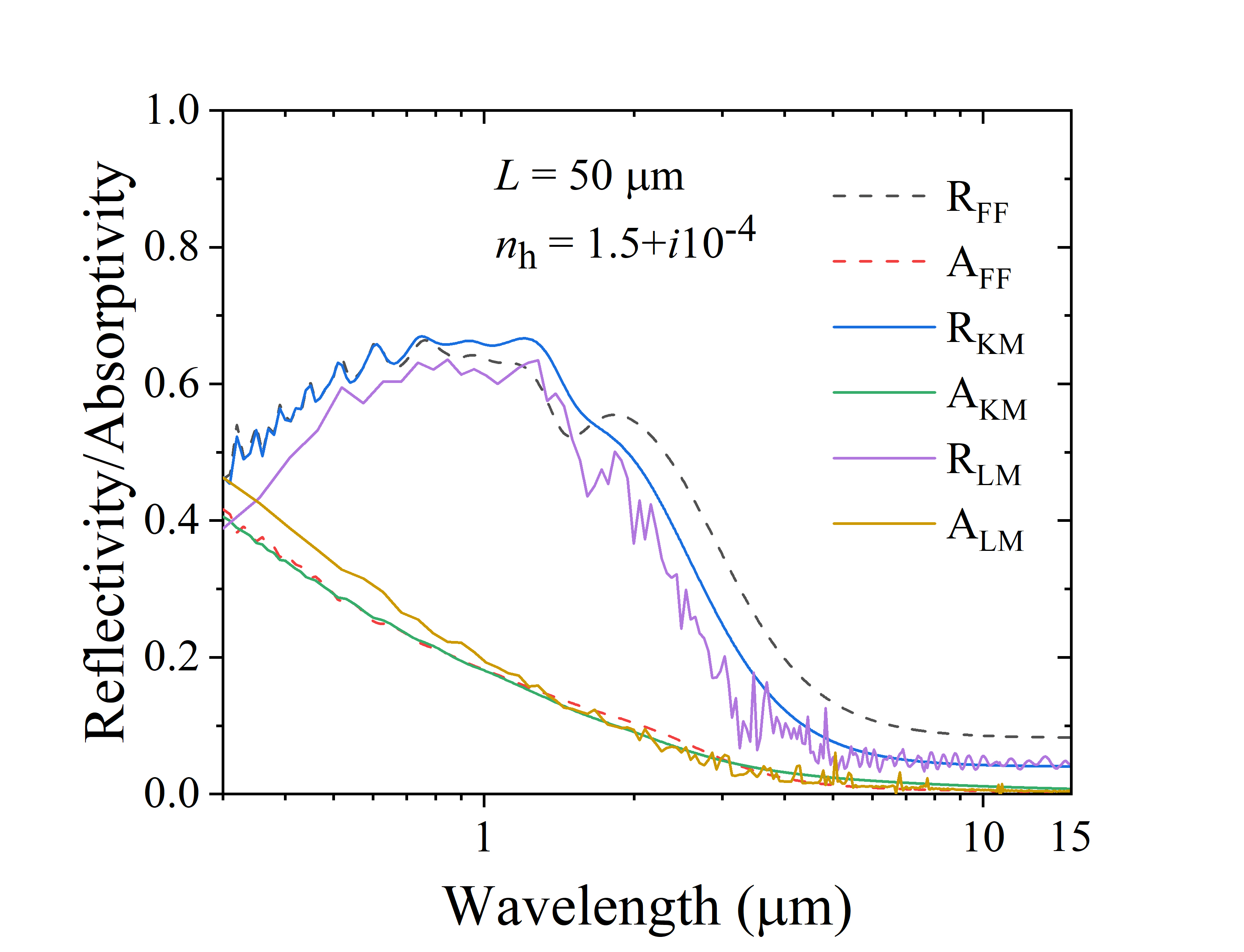}
   \caption{}
    \label{fig:AB_sr2}
     \end{subfigure}
     \hfill
    \begin{subfigure}[b]{0.49\textwidth}
    \centering
       \includegraphics[width=7cm]{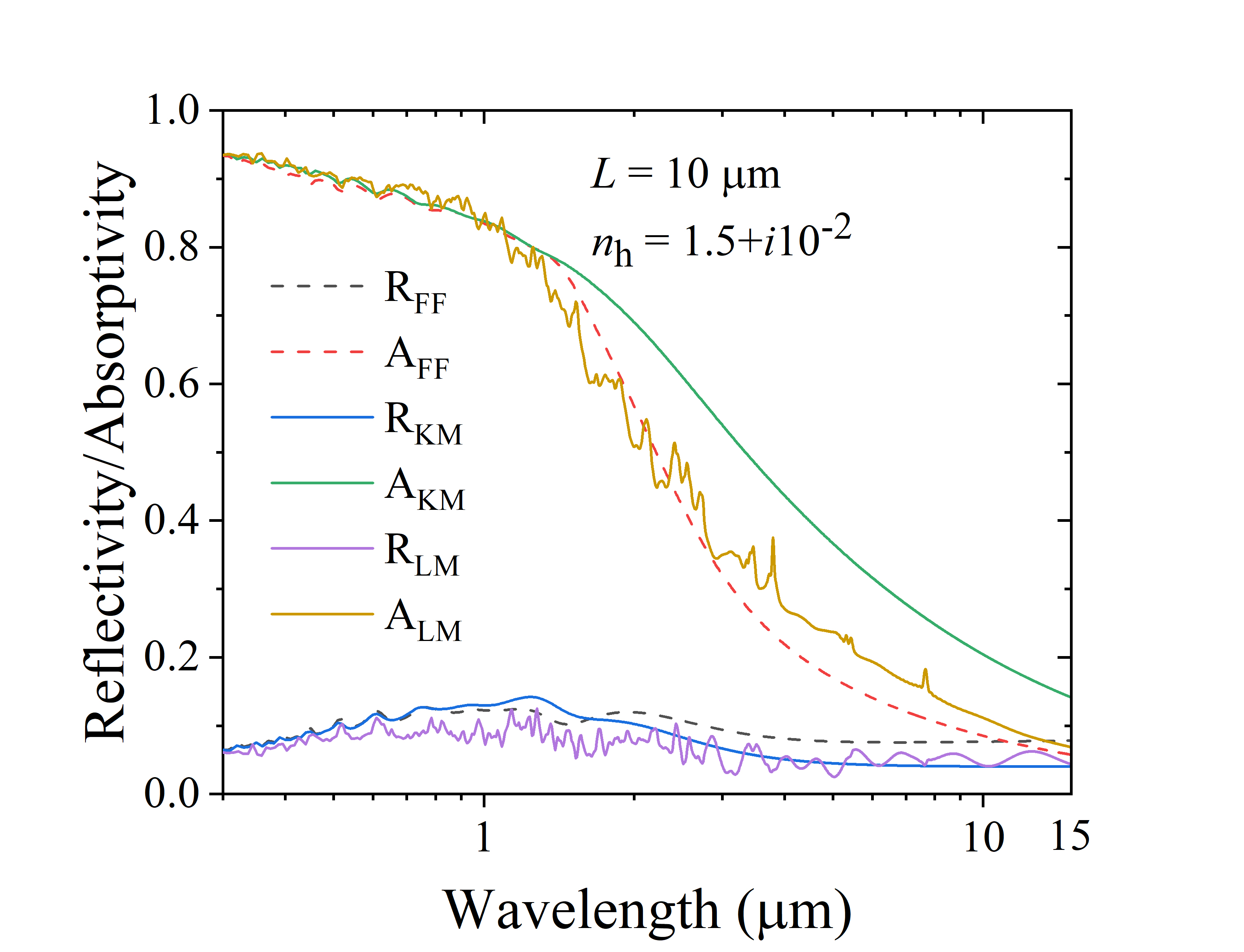} 
    \caption{}
    \label{fig:AB_sr5}
     \end{subfigure}
     \hfill
     \begin{subfigure}[b]{0.49\textwidth}
    \centering
        \includegraphics[width=7cm]{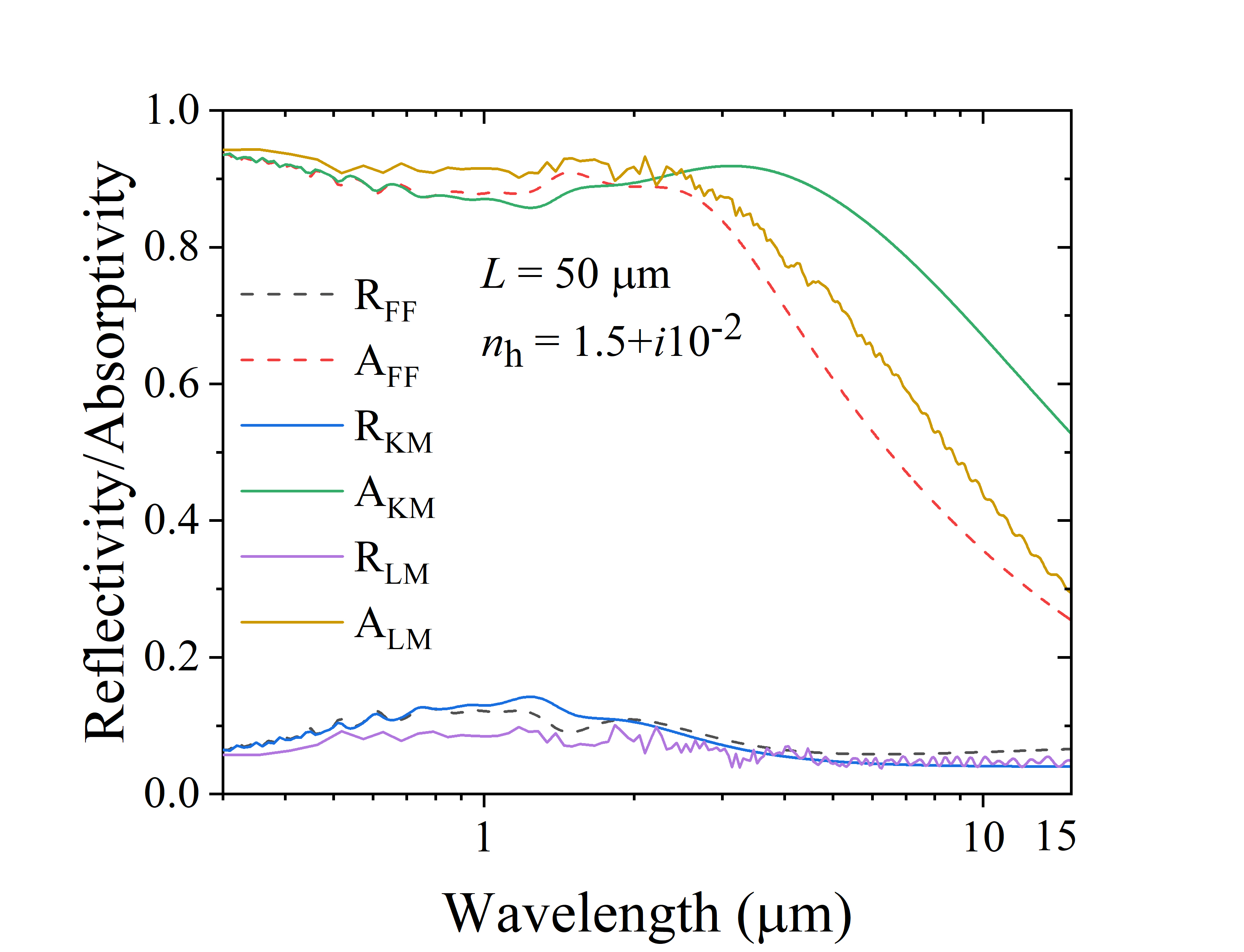}
    \caption{}
    \label{fig:AB_sr6}
     \end{subfigure}
    \caption{ Reflectivity and absorptivity spectra for $n_{\text{p}}=2.5$, $r=0.25~\mu\text{m}$,  $f=0.05$, and   (a) $n_{\text{h}}=1.5 + i10^{-4},$ $L = 10~\mu\text{m}$; (b)  $n_{\text{h}}=1.5+i10^{-4},$ $L = 50~\mu\text{m}$;  (c)  $n_{\text{h}}=1.5+i10^{-2},$ $L = 10~\mu\text{m}$; (d) $n_{\text{h}}=1.5+i10^{-2},$ $L = 50~\mu\text{m}$. 
    Here, FF stands for four flux, KM for Kubelka Munk, and LM for Lumerical.}
    \label{fig:AB_sr1_2}
\end{figure}

In the presence of absorbing host media, the expressions for $\gamma$ and $A$  in Eqs. \ref{eq:KM_R} and \ref{eq:KM_T} needs to be modified to account for absorption in the matrix  \cite{Etherden2004,Gali2017a} as:
$\gamma = \sqrt{k'/(  k' + s')}$, and $A = \sqrt{2k'(2k'+s')}$ where,
$
    k' = k+(1-f)\alpha.
$
Here, $\alpha = 4\pi n''_\text{h}/\lambda$, with $n''_\text{h}$ being the imaginary part of refractive index of the matrix, and $\lambda$  the wavelength in vacuum. 
In addition, expressions for $Q_{\text{sca}}$ and $Q_{\text{abs}}$ in Eq. \ref{eq:Qsca_Qabs} needs to be modified as shown by Mischenko et al. \cite{Mishchenko2018}. 
Figure \ref{fig:AB_sr1} and \ref{fig:AB_sr2} show the  comparison between KM, FF, and FDTD results for the case when host medium is weakly absorbing with $n_h'' = 10^{-4}$ and Fig. \ref{fig:AB_sr5} and \ref{fig:AB_sr6} show the corresponding comparison when it is more strongly absorbing with $n_h'' = 10^{-2}$. 
In the presence of weakly absorbing matrix and for smaller thickness of the coating FF is again observed to deviate significantly from the FDTD simulations.  As absorption increases we observe significant deviation from FDTD results in both FF and KM theories particularly for the higher wavelengths.

\subsection{Comparison of predictions from  KM, FF theories and FDTD solver in the independent scattering regime for polydisperse inclusions with and without absorption in particles}
\label{sec:poly}

\begin{figure}[ht]
     \centering
     \begin{subfigure}[b]{0.49\textwidth}
    \centering
        \includegraphics[width=7cm]{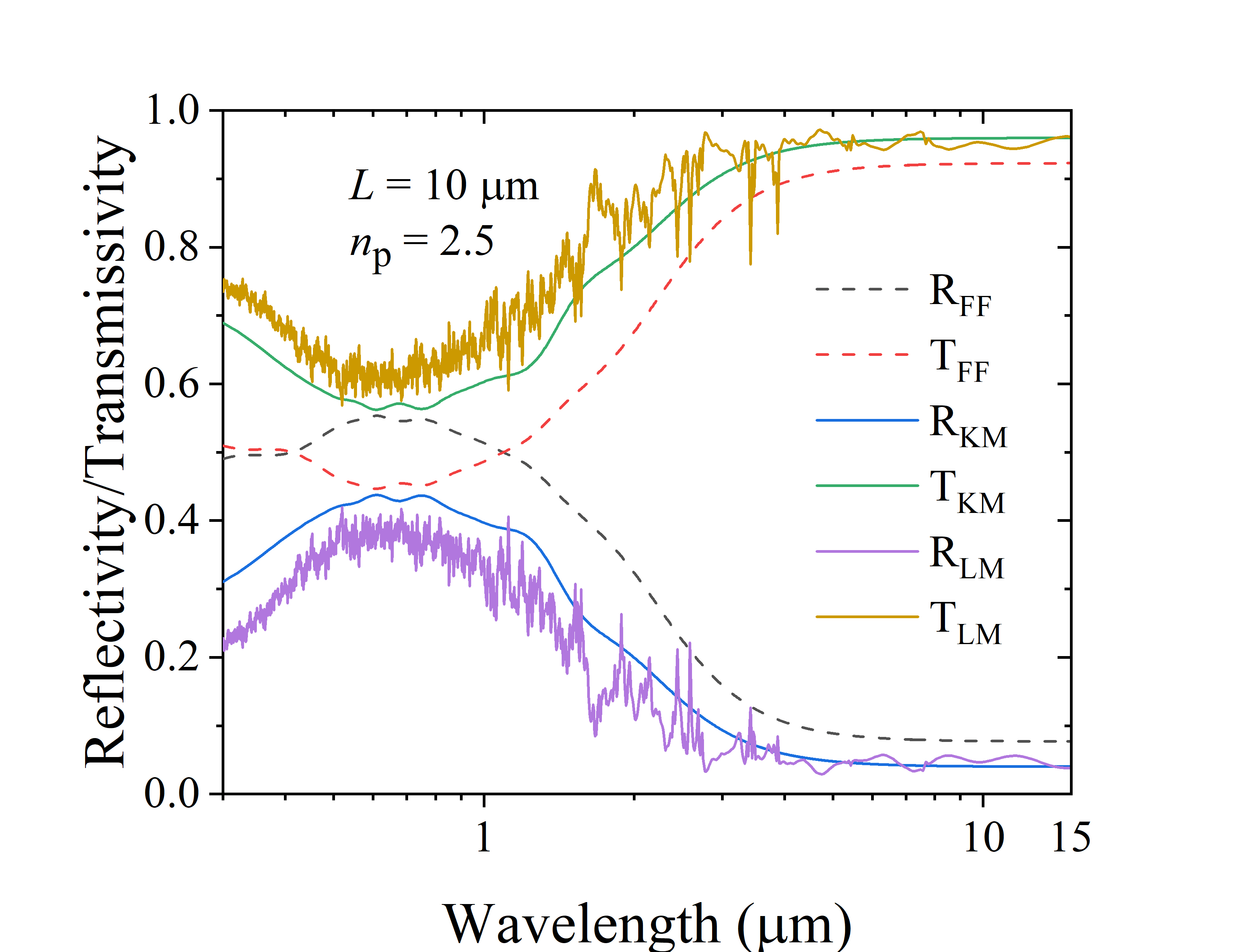}
    \caption{}
    \label{fig:poly_sr1}
     \end{subfigure}
     \hfill
     \begin{subfigure}[b]{0.49\textwidth}
    \centering
        \includegraphics[width=7cm]{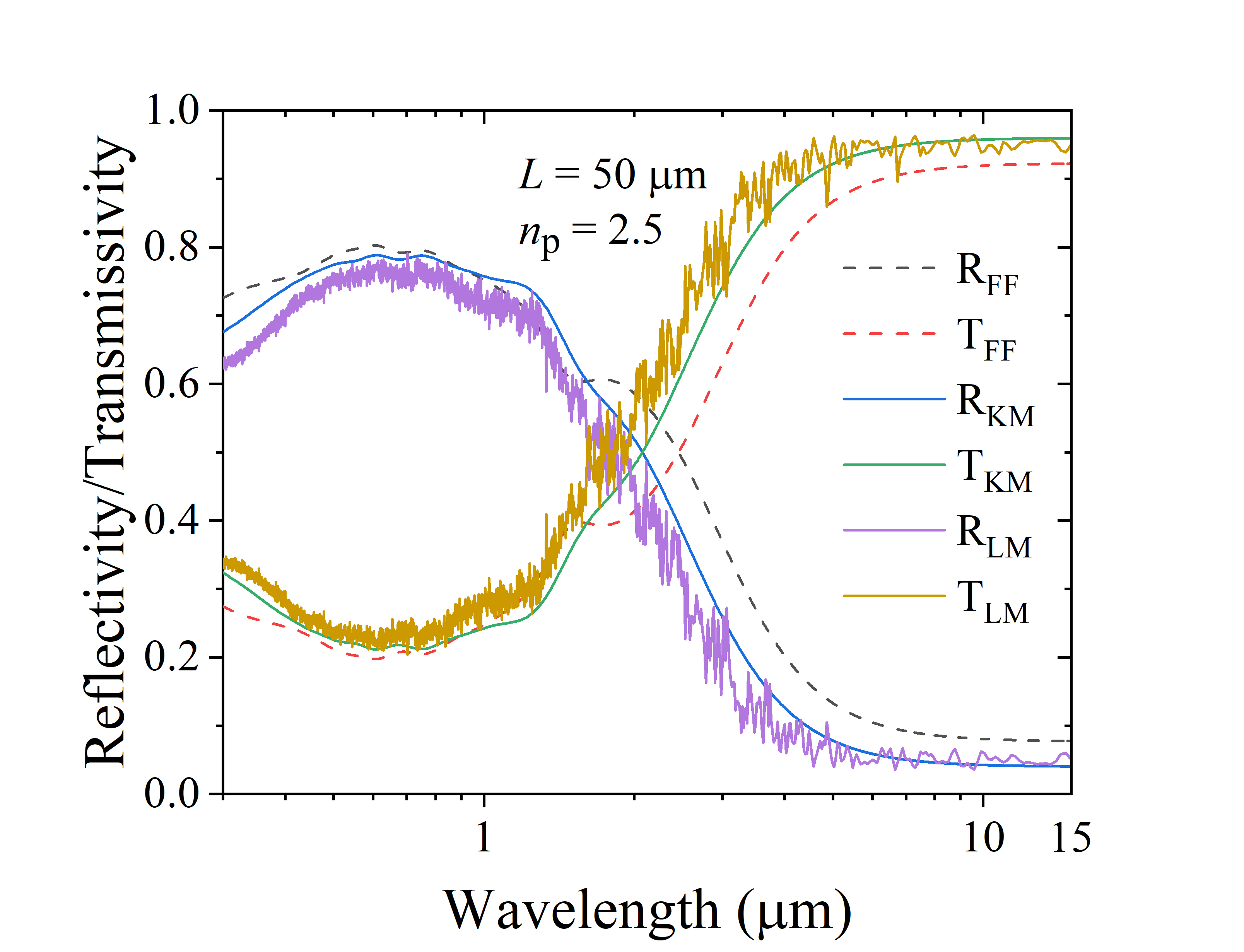}
    \caption{}
    \label{fig:poly_sr2}
     \end{subfigure}
     \hfill
      \begin{subfigure}[b]{0.49\textwidth}
    \centering
        \includegraphics[width=7cm]{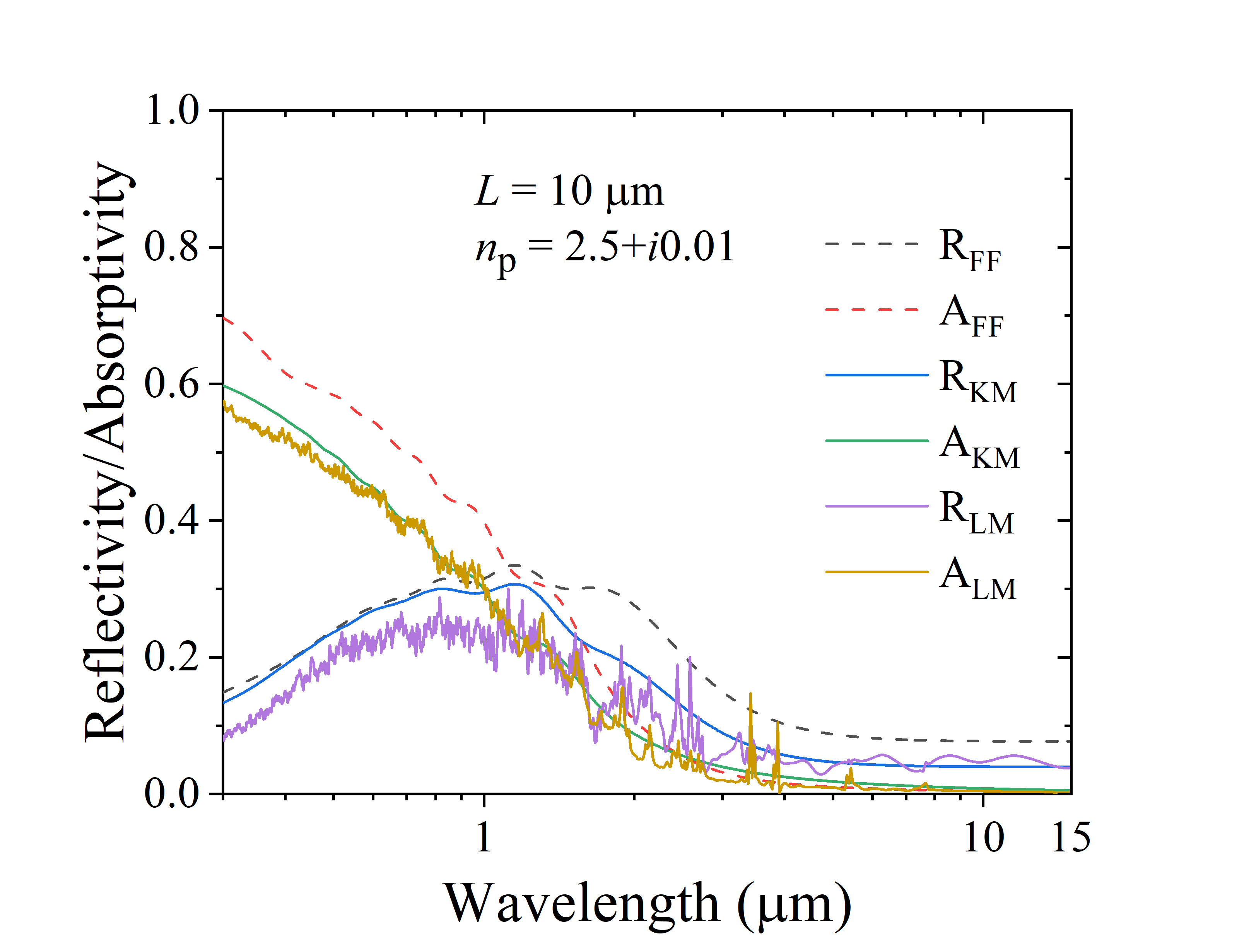}
    \caption{}
    \label{fig:poly_sr3}
     \end{subfigure}
     \hfill
     \begin{subfigure}[b]{0.49\textwidth}
    \centering
        \includegraphics[width=7cm]{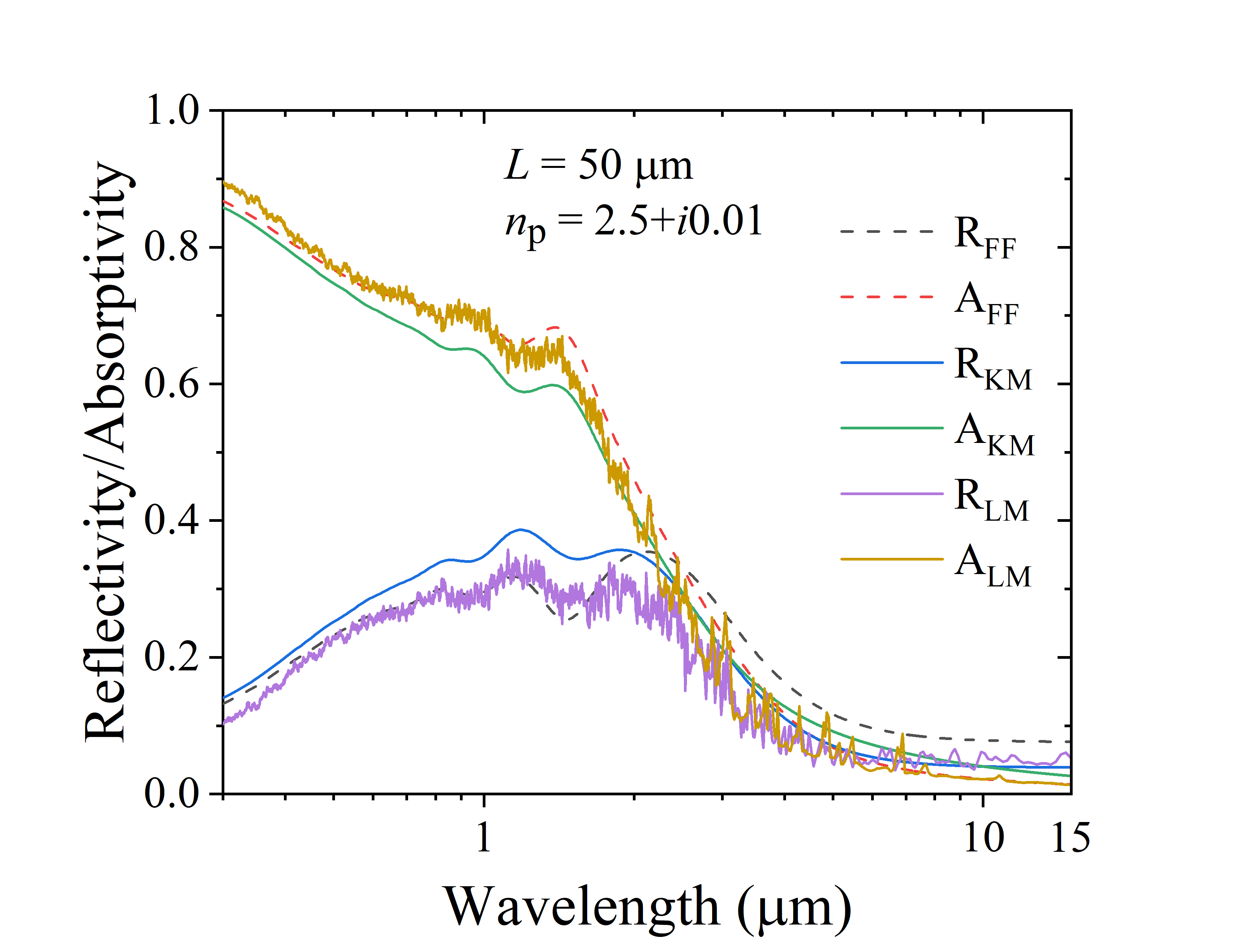}
    \caption{}
    \label{fig:poly_sr4}
     \end{subfigure}
    \caption{Reflectivity and transmissivity for $n_{\text{h}}=1.5,$ $r=0.25~\mu\text{m},$ $\sigma=0.016,$ $f=0.05$ and (a) $n_{\text{p}}=2.5,$ $L=10~\mu\text{m}$; (b) $n_{\text{p}}=2.5,$ $L=50~\mu\text{m}$. Reflectivity and absorptivity for $n_{\text{h}}=1.5,$ $r=0.25~\mu\text{m},$ $\sigma=0.016,$ $f=0.05$ and (c) $n_{\text{p}}=2.5+0.01i,$ $L=10~\mu\text{m}$; (d) $n_{\text{p}}=2.5+0.01i,$ $L=50~\mu\text{m}$. Here, FF stands for four flux, KM for Kubelka Munk, and LM for Lumerical.}
    \label{fig:poly_sr1_2}
\end{figure}
In this section we explore the predictive capability of KM and FF theories for polydisperse medium which consists of randomly positioned particles with different sized radius.
The study is motivated from the observation that synthesis of nanoparticles via various methods such as sol-gel \cite{bokov2021nanomaterial}, microemulsion \cite{malik2012microemulsion}, hydrothermal \cite{gan2020hydrothermal},
results in a polydisperse size distribution of particles. 
%
%
Moreover, some recent studies \cite{Peoples2019,Li2021baso4}  have also deliberately adopted coatings with different size distribution of particles to make use of the property of size-dependent scattering of particles to obtain wavelength-selective coatings. 
Such a particulate medium can be analyzed by considering the particles to be distributed about a mean radius $r$ with  standard deviation $\sigma$, with 
the expressions for $s$ and $k$ to be used in Eqs. \ref{eq:Qsca_Qabs}  got by summing over the respective coefficients for individual particle volume fractions $f_i$ \cite{Gali2017a} as:
\begin{equation}
    s = \sum_{i=1}^{N} s_i;  \,\,\,\,  k = \sum_{i=1}^{N} k_i
    \label{eq:poly_S}
\end{equation}
%
where $s_i$ and $k_i$ are the Mie scattering and absorption coefficients respectively of the particle with fill fraction $f_i$. Equation (\ref{eq:poly_S})  can also be used to calculate $s$ and $k$ when there are two or more type of particles present in the matrix (with different refractive index).
%
For demonstration we consider a Gaussian distribution of spherical particles about mean radius $r = 0.25~\mu m$ with  standard deviation $\sigma = 0.016~ \mu m$ with and without absorption. The particle size distribution curve has been shown in Fig. S2.  
%
Figure \ref{fig:poly_sr1}  and \ref{fig:poly_sr2} show the  comparison between KM, FF, and FDTD results for the case when particles are non-absorbing  and Fig. \ref{fig:poly_sr3} and \ref{fig:poly_sr4} show the corresponding comparison when particles are absorbing with $n_p'' = 10^{-2}$.
Other parameter values are retained as for the case of monodisperse particulate coating. 
%
The observations follow the trend seen for the case of monodisperse coating with significant deviations observed in the predictions of FF method for lower values of thicknesses of coatings and when particles are nonabsorbing. For larger thicknesses of the coating and in the presence of absorption, both FF and KM are observed to predict the optical properties with reasonable accuracy across the spectrum. 

\subsection{Comparison of predictions from  KM, FF and FDTD solvers in the dependent scattering regime}
\label{sec:dep_scat}

\begin{figure}[ht]
     \centering
     \includegraphics[width=7cm]{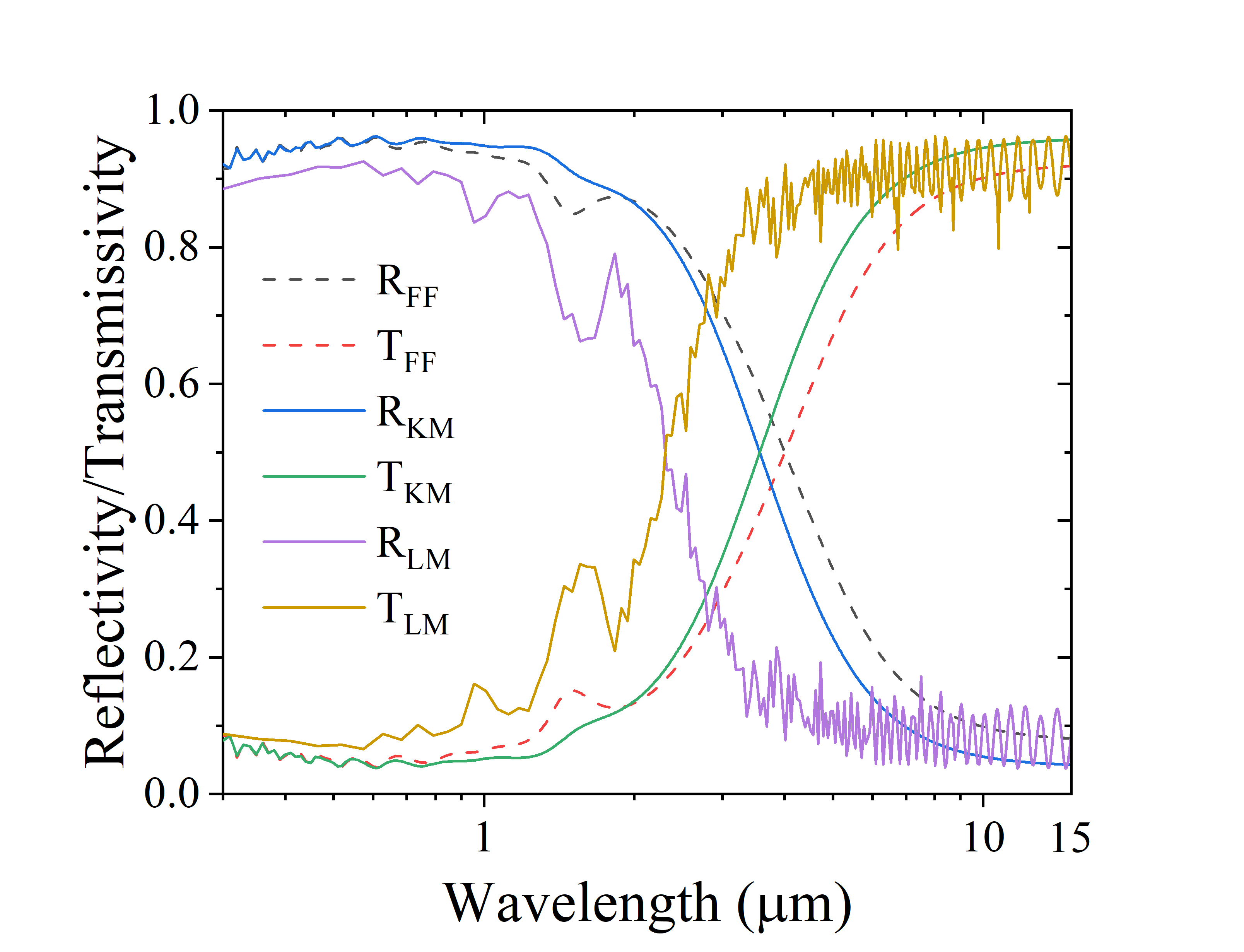}
    \caption{Reflectivity and transmissivity for $n_{\text{p}}=2.5,$ $n_{\text{h}}=1.5,$ $r=0.25~\mu\text{m},$ $f=0.3,$ and $L=50~\mu\text{m}$. Here, FF stands for four flux, KM for Kubelka Munk, and LM for Lumerical.
    }
    \label{fig:dep_scat}
\end{figure}

So far we have analyzed for the situations where the fill fraction $f$ of particles in the composite is small enough so that the particles can be assumed to independently scatter from one another.   However, as the fill fraction of particles increases, there will be a transition to  dependent-scattering regime where both the near-field interaction between the particles as well as far-field interference between the scattered field of individual particles have a significant impact on the overall properties of the coating. Hotel \cite{Hottel1971} empirically determined this transition to occur when $f> 0.27$ and $\overline{d}/\lambda > 0.3$ where $\overline{d}$ is the mean inter-particle spacing and $\lambda$ is the wavelength. Several coatings reported in literature \cite{Bao2017,Atiganyanun2018,Mandal2018,Li2020CaCO3,Li2021baso4} have fill fractions in the range 0.1-0.6 where such effects cannot be neglected. We thus explore here the predictive capability of FF and KM theories  for such coatings by considering a monodisperse distribution of particles with increased fill fraction $f = 0.3$ while retaining other parameter values to be same as that included for Fig. \ref{fig:NAB_sr2}. This comparison is shown in Fig. \ref{fig:dep_scat}, where we observe that the predictions from both FF and KM theories deviate significantly from FDTD simulations across the spectra, and thus cannot be relied on for predicting optical properties of such coatings.

\begin{table}[ht]
\centering
\caption{Weighted-average reflectivity in solar spectrum  $R_{\text{solar,KM}}$ ($R_{\text{solar,FF}}$) and emissivity in IR spectrum $\epsilon_{\text{IR,KM}}$ ($\epsilon_{\text{IR,FF}}$) calculated using KM (FF) theory. Values in brackets denote deviation of prediction from FDTD results.}
\label{Table:comp1}
\begin{tabular}{llllll}
\hline
\textbf{Sr. no.}                                                  & \textbf{Fig. no.} & \textbf{$R_{\text{solar,KM}}$} & \textbf{$\epsilon_{\text{IR,KM}}$} & \textbf{$R_{\text{solar,FF}}$} & \textbf{$\epsilon_{\text{IR,FF}}$}  \\ \hline
1                                                   &    \ref{fig:NAB_sr1}                               & 0.391 (21.4\%)    & -                                & 0.509 (58.1\%)    & -                                                               \\
\begin{tabular}[c]{@{}l@{}}2\end{tabular}        &     \ref{fig:NAB_sr2}                              & 0.745 (0.67\%)     & -                                & 0.747 (0.4\%)     & -                                                                \\
\begin{tabular}[c]{@{}l@{}}3\end{tabular}   &   \ref{fig:NAB_sr5}                                & 0.076 (13.4\%)      & 0.08  (95\%)                         & 0.081 (20.9\%)     & 0.043 (4.87\%)                                                      \\
\begin{tabular}[c]{@{}l@{}}4\end{tabular}   &   \ref{fig:NAB_sr6}                                & 0.078 (18.2\%)     & 0.331 (70.6\%)                         & 0.079 (19.7\%)    & 0.197 (1.55\%)                                                   \\
\begin{tabular}[c]{@{}l@{}}5\end{tabular} &    \ref{fig:AB_sr1}                               & 0.376 (18.6\%)    & -                         & 0.483 (52.4\%)    & -                                                  \\
\begin{tabular}[c]{@{}l@{}}6\end{tabular} &     \ref{fig:AB_sr2}                              & 0.619 (8.22\%)    & 0.012 (100\%)                          & 0.612 (6.99\%)     & 0.005 (16.67\%)                                                    \\
\begin{tabular}[c]{@{}l@{}}7\end{tabular}   &   \ref{fig:AB_sr5}                                & 0.112 (30.2\%)     & 0.205 (83.0\%)                         & 0.110 (27.9\%)    & 0.086 (23.2\%)                                                     \\
\begin{tabular}[c]{@{}l@{}}8\end{tabular}   &    \ref{fig:AB_sr6}                               & 0.112 (36.6\%)     & 0.661  (51.3\%)                         & 0.108 (31.7\%)    & 0.357 (18.3\%)                            
\\
\begin{tabular}[c]{@{}l@{}}9\end{tabular}        &    \ref{fig:poly_sr1}                               & 0.392 (19.1\%)    & -                                & 0.510 (55.0\%)    & -                                                               \\
\begin{tabular}[c]{@{}l@{}}10\end{tabular}        &   \ref{fig:poly_sr2}                                & 0.746 (4.63\%)    & -                                & 0.755 (5.89\%)    & -                                                              \\
\begin{tabular}[c]{@{}l@{}}11\end{tabular}  &   \ref{fig:poly_sr3}                                & 0.260 (25.0\%)     & 0.008 (60\%)                           & 0.279 (34.1\%)    & 0.004 (20\%)                                                      \\
\begin{tabular}[c]{@{}l@{}}12\end{tabular}  &    \ref{fig:poly_sr4}                               & 0.304 (16.5\%)    & 0.041 (78.3\%)                         & 0.268 (2.68\%)    & 0.023  (0.01\%)                                                \\
\begin{tabular}[c]{@{}l@{}}13\end{tabular}  &    \ref{fig:dep_scat}   & 0.944 (8.13\%)     & -                         & 0.935 (6.98\%)   &-                     \\    \hline
 &     &       &          &            &                      
\end{tabular}
\end{table}

A comparison between the weighted average of the optical properties across the spectra 
 as predicted by KM and FF theories for the different cases considered so far has been tabulated in Table \ref{Table:comp1}. The weighted averages are calculated as: $R_{\text{solar}} = \int I_{\text{AM1.5}}(\lambda) R(\lambda) d \lambda / \int I_{\text{AM1.5}}(\lambda) \, d\lambda$, where $I_{\text{AM1.5}}(\lambda)$ is the spectral solar irradiance \cite{AM1.5} and $\epsilon_{\text{IR}} =\int I_{\text{BB}}(\lambda) \epsilon(\lambda) d \lambda / \int I_{\text{BB}}(\lambda) \, d\lambda$, where $I_{\text{BB}}(\lambda)$ is the black body irradiance. For the relevant applications in consideration for this study i.e. coatings suitable for radiative cooling application and for use in solar thermal absorber plates,  the reflection over the solar spectrum i.e. over wavelength range $0.3-3\, \mu m$ and emissivity over the infra-red spectrum i.e. over wavelength range $5 - 15\, \mu m$ is of primary importance, and the weighted average over this spectral range 
 is reported in Table \ref{Table:comp1} along with  the deviation from FDTD simulations expressed in \% error in brackets. 


\section{Semi-analytical method}
\label{sec:semi_anyl}

\begin{figure}[ht]
     \centering
     \includegraphics[width=7cm]{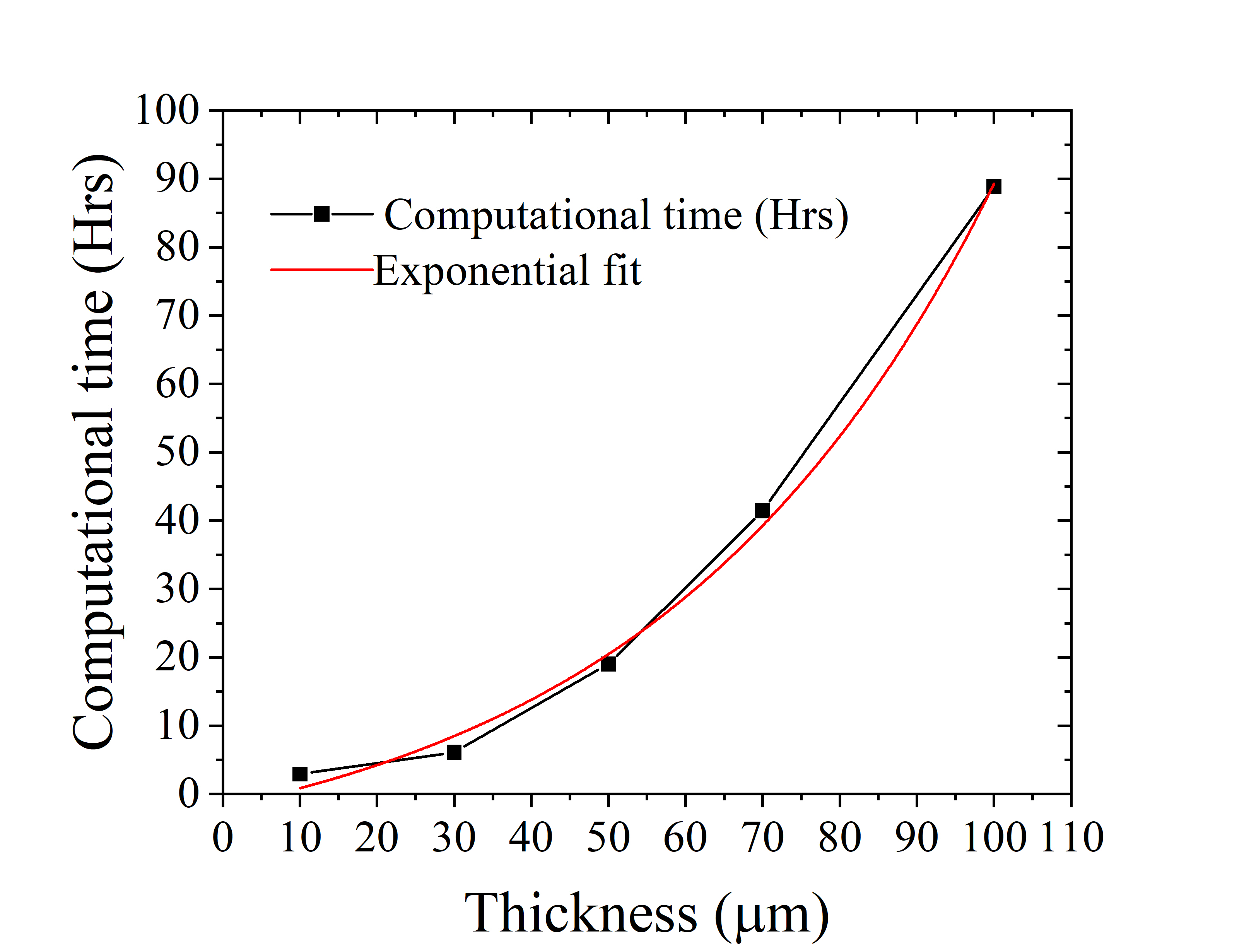}
    \caption{Comparison of computational time as a function of thickness of the disordered media coating. Simulations are carried out in ANSYS Lumerical using an eight-core Intel Xeon workstation for the configuration: $n_p= 2.5$, $n_h= 1.5$, $r = 0.25~\mu$m, $f=  0.05$, with mesh size 30~nm. Auto shutoff level (simulation termination criteria) is set at 10$^{-3}$. }
    \label{fig:comptime}
\end{figure}

The comparison with FDTD simulations shown in Section \ref{sec:analytical} demonstrate the failure of KM and FF analytical methods in configurations where dependent scattering is not negligible and when matrix/particles are absorbing. This failure can be attributed to the actual scattering and absorption coefficients of these coatings diverging from the values calculated using Mie scattering coefficients of the individual particles. At present no single analytical technique exists that can correctly predict the optical properties of particulate media in the presence of dependent scattering effects as well as correctly account for the absorption in matrix/particles. 
One can then resort to using exact numerical solvers to accurately estimate the optical properties of the coating in such cases. However, as Fig. \ref{fig:comptime} shows, the computational time required to simulate such structures increases exponentially with thickness of the coating. For coatings of thickness in the range 100-500 microns which are currently being adopted in literature for the radiative cooling application \cite{Huang2017,Li2020CaCO3,Li2021baso4,Zhang2021Zro2,Mishra2021} the design time is clearly prohibitive. 
In such cases it becomes imperative to develop alternate techniques which can combine the accuracy power of exact FDTD solvers with the simplicity and minimal computational requirements of the analytical techniques. Particularly when multiple parameters are involved in design - such as that observed for disordered media - such a method will prove to be useful in reducing the design time to find the optimum combination of parameters necessary to obtain the required optical properties of the coating.  
In order to obtain a better estimate for the absorption and scattering coefficients of such media where dependent scattering effects are non-negligible, researchers have previously \cite{Molenaar1999,Levinson2005,Barrios2013,Wang2018} relied on experimental measurements of the optical properties of a fabricated coating  and then using the KM theory results from Section \ref{sec:analytical}  to extract the required coefficients. Instead of relying on experimental measurements which is not always feasible especially at the initial state of design,  we modify this technique and instead propose the 
 following two-step semi-analytical method to estimate the optical properties of  random media of thickness $L$ when usage of exact numerical solvers to simulate the properties of such a thick coating is prohibitive.
\begin{itemize}
    \item 
\textbf{Step 1:} Use a numerical solver to obtain the optical properties $R$ and $T$ of a similar coating but with much smaller thickness $t_s \ll L$  and extract the $\gamma$ and $A$ parameters by inverting Eq. \ref{eq:KM_R} and \ref{eq:KM_T}.  Care must be taken at this step to ensure that the configuration set up in the solver considers incident light to be in the same medium as the  index of matrix i.e., $n_{\text{surr}} = n_\text{h}$ in order to ensure that reflection from surfaces and substrates are not included in this step. In case the host matrix is absorbing then only the real part is considered i.e., $n_{\text{surr}} = \text{Re}(n_\text{h})$.
Care must also be taken to ensure that when scattering efficiency of the particles is high, the value of $t_s$ should be chosen such that $t_s \gg l_s$ where $l_s \approx 1/(N \sigma_s)$ is the scattering mean-free path with $N$ being the particle number density and $\sigma_s$ the scattering cross section.   At the other limit when scattering efficiency is low the optical properties of the coating are primarily determined from surface reflection and transmission which are accounted for in step 2. Thus the choice of $t_s$ is determined from the scattering mean-free path calculated in the high-scattering regime. 
\item
\textbf{Step 2}:
From the  $\gamma$ and $A$ parameters extracted from step 1 use the analytical expressions from KM theory i.e. Eqs. \ref{eq:KM_R}, \ref{eq:KM_T}, \ref{eq:KM_Rsurf} and \ref{eq:KM_Rb} to predict the optical properties of the coating of the required thickness $L \gg t_s$. Specular reflection at the surfaces as well as at the substrate are accounted for here. 
\end{itemize}
A more elaborate procedure, along with details of a supporting convergence test which may need to be incorporated in some cases to arrive at the value of thickness $t_s$ is included in  Section S4 of supplementary.

\begin{figure}[ht]
     \centering
     \begin{subfigure}[b]{0.49\textwidth}
    \centering
        \includegraphics[width=7cm]{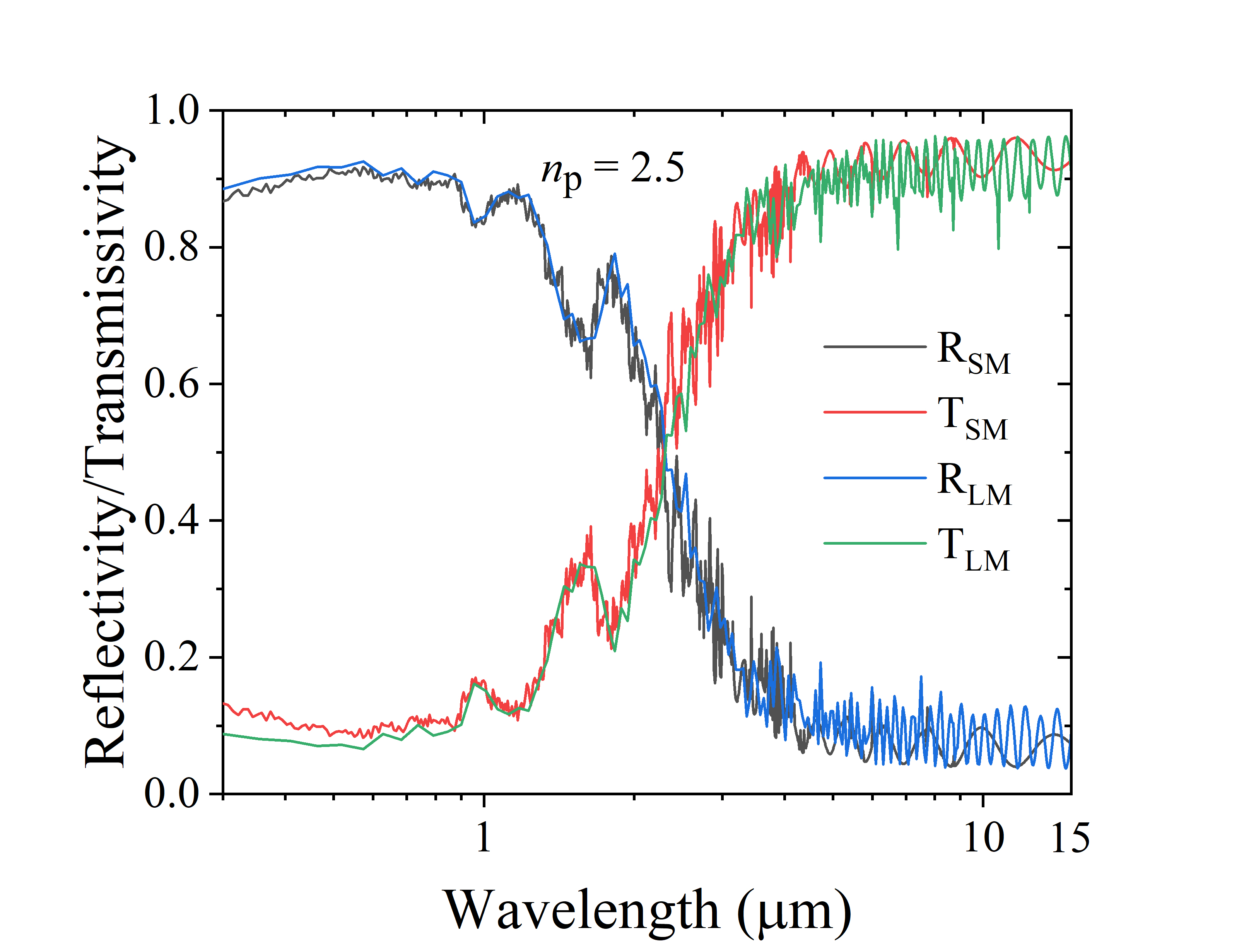}
    \caption{}
    \label{fig:NAB_sr4}
     \end{subfigure}
     \hfill
     \begin{subfigure}[b]{0.49\textwidth}
    \centering
        \includegraphics[width=7cm]{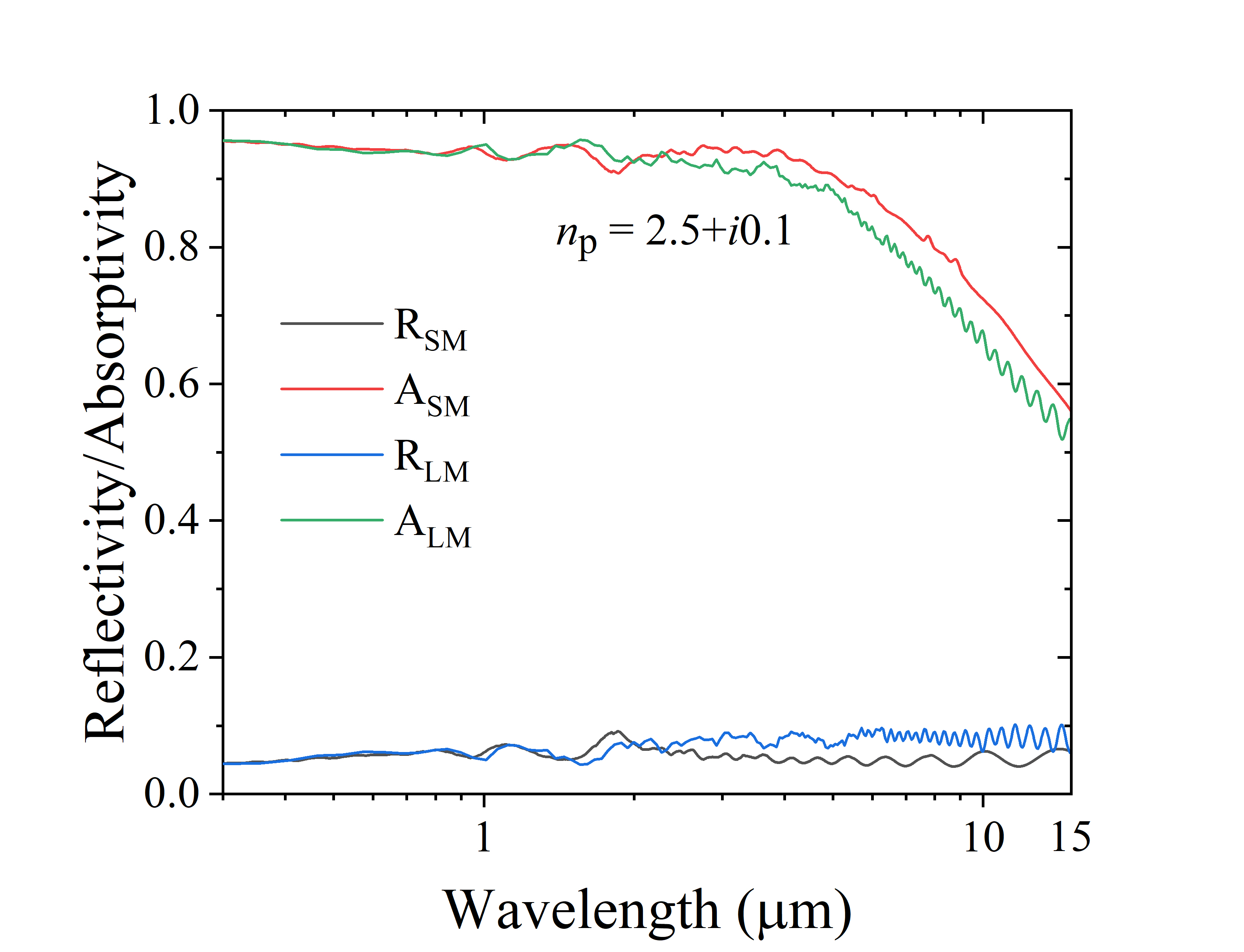}
   \caption{}
    \label{fig:NAB_sr8}
     \end{subfigure}
    \caption{(a) Reflectivity and transmissivity for $n_{\text{p}}=2.5$, $n_{\text{h}}=1.5,$ $r=0.25~\mu\text{m},$ $f=0.3,$ and $L=50~\mu\text{m}$; (b) Reflectivity and absorptivity for $n_{\text{p}}=2.5+0.1i$, $n_{\text{h}}=1.5,$ $r=0.25~\mu\text{m},$ $f=0.3,$ and $L=50~\mu\text{m}$. Here, SM stands for semi-analytical method and LM for Lumerical.}
    \label{fig:NAB_sr4_8}
\end{figure}
\begin{figure}[h!]
     \begin{subfigure}[b]{0.49\textwidth}
    \centering
        \includegraphics[width=7cm]{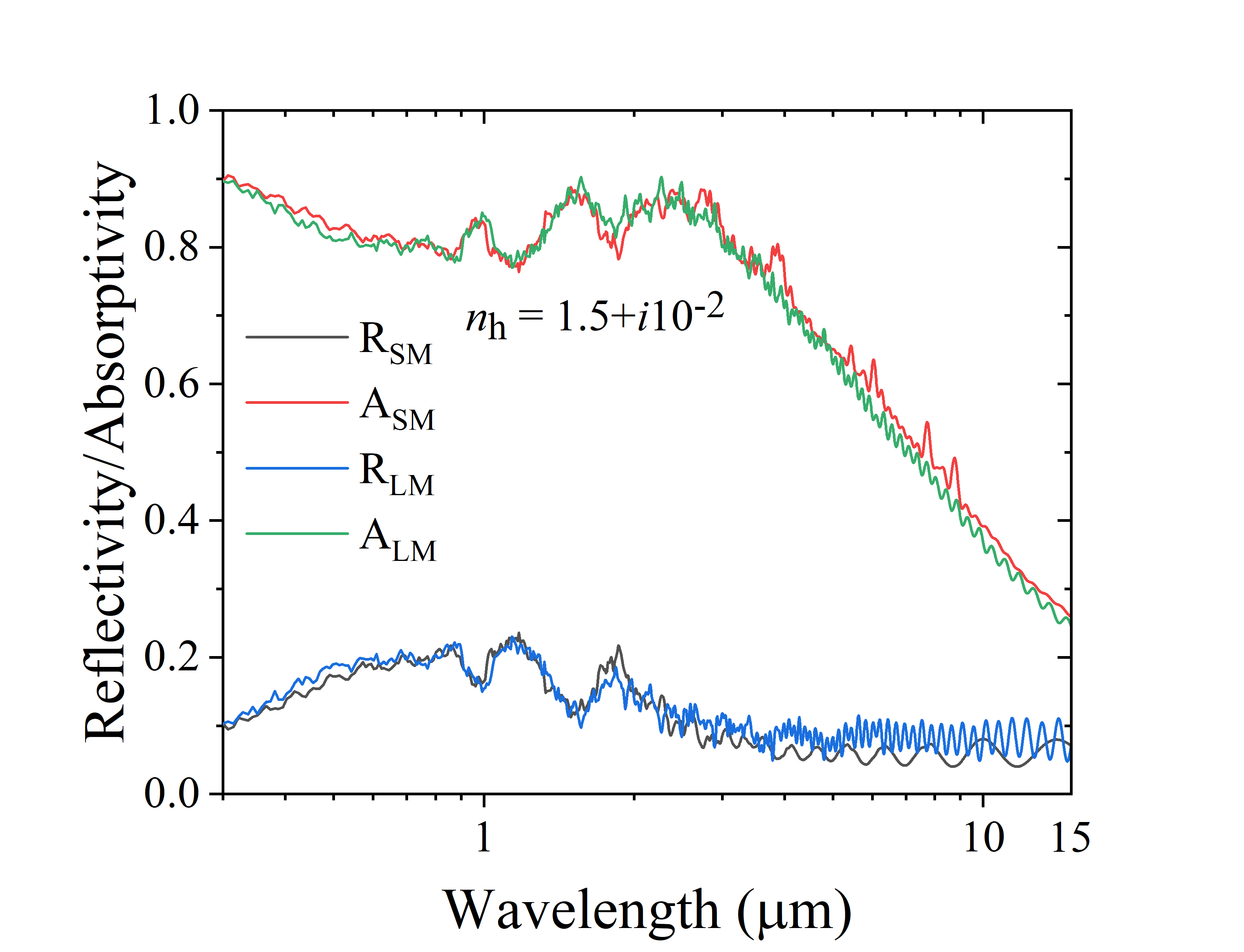}
    \caption{}
    \label{fig:AB_sr8}
     \end{subfigure}
     \hfill
     \begin{subfigure}[b]{0.49\textwidth}
    \centering
        \includegraphics[width=7cm]{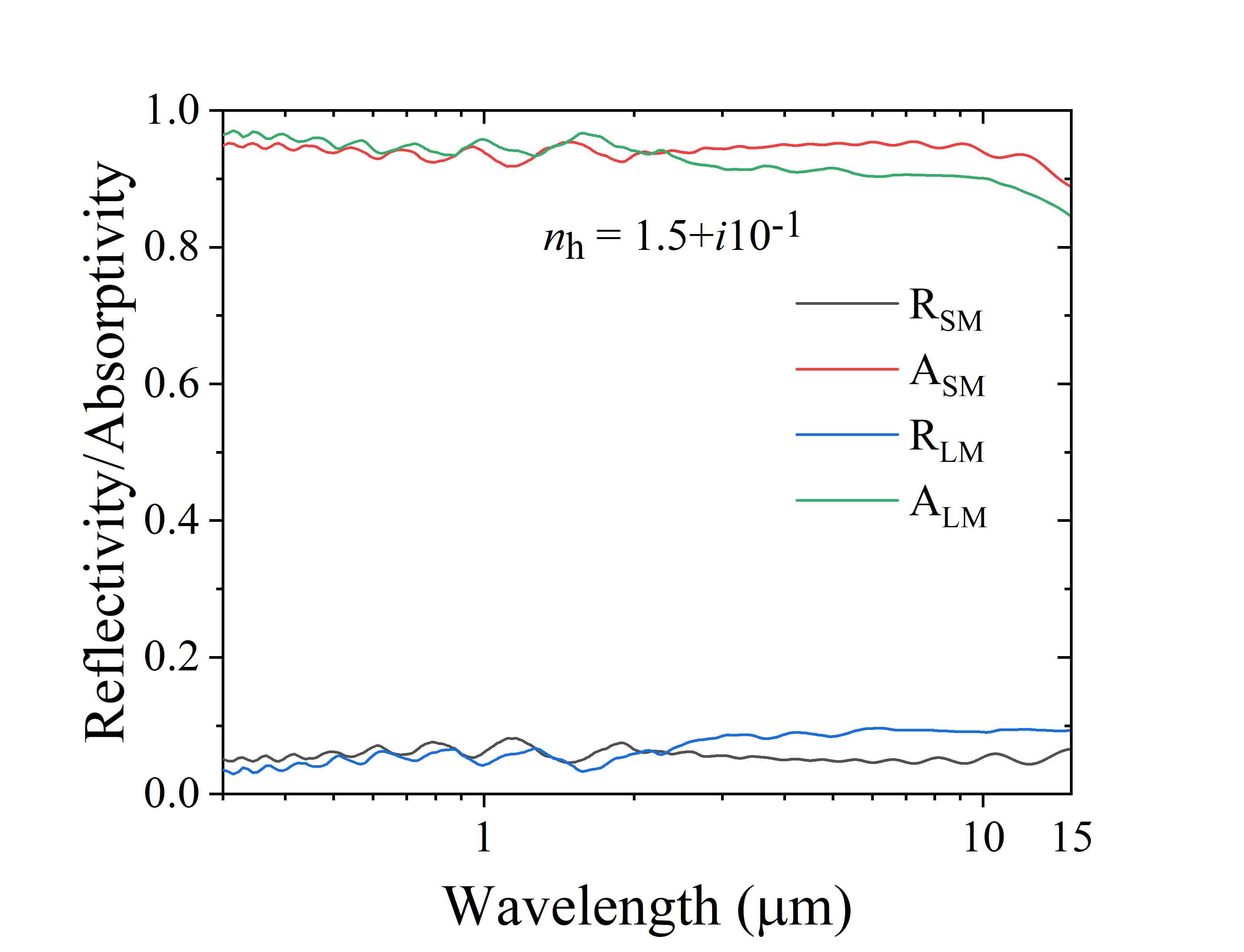}
    \caption{}
    \label{fig:AB_sr12}
     \end{subfigure}
    \caption{Reflectivity and absorptivity for $n_{\text{p}}=2.5,$ $r=0.25~\mu\text{m},$  $f=0.3,$ $L=50~\mu\text{m},$ and (a) $n_{\text{h}}=1.5+i10^{-2}$; (b) $n_{\text{h}}=1.5+i10^{-1}$. Here, SM stands for semi-analytical method and LM for Lumerical.}
    \label{fig:AB_sr4_8_12}
\end{figure}

We now apply this technique for the cases considered in Section \ref{sec:analytical} where the predictions from analytical methods deviated significantly from those of FDTD solver, such as for the dependent scattering regime, as well as when the absorption in the particles/host matrix is significant.  Fig. \ref{fig:NAB_sr4_8} (Fig. \ref{fig:AB_sr4_8_12}) shows the comparison between the predictions from the semi-analytical technique and from FDTD simulations when absorption in particles (host matrix) is varied. In both these cases the semi-analytical technique uses the results of exact FDTD simulations of a $10~\mu$m thick coating to predict the optical properties of a larger  $50~\mu$m thick coating. A volume fill fraction $f = 0.3$ is maintained in both these cases where dependent scattering effects are known to be dominant, while keeping other parameter values same as that analysed for the monodisperse case of Sec. \ref{sec:mono}.
For these cases we observe a close match in the predictions of the semi-analytical method with the FDTD results over the entire spectrum, with only a slight deviation observed for the higher wavelengths when absorption is high. The weighted-average reflectivity  of the coating for solar spectrum, and emissivity over the infra-red spectrum for the cases considered in Figs. \ref{fig:NAB_sr4_8} and \ref{fig:AB_sr4_8_12} are listed in Table \ref{Table:comp2} along with  the deviation from FDTD simulations expressed in \% error in brackets. Particularly illustrative of the effectiveness of the semi-analytical technique is the reduction in error (1.03 \% in Sr. no. 1) as compared to those obtained from analytical techniques and reported in Table \ref{Table:comp1} ( 8.13 \% using KM theory and 6.98 \% using FF theory in Sr. no. 13) for the configuration: $n_{\text{p}}=2.5$, $n_{\text{h}}=1.5,$ $r=0.25~\mu\text{m},$ $f=0.3,$ and $L=50~\mu\text{m}$ where dependent scattering is expected to be dominant. 

\begin{table}[ht]
\centering
\caption{Weighted-average reflectivity in solar spectrum  $R_{\text{solar,SM}}$, and emissivity in IR spectrum $\epsilon_{\text{IR,SM}}$ calculated using the semi-analytical technique. Values in brackets denote deviation of prediction from FDTD results.} 
\label{Table:comp2}
\begin{tabular}{llll}
\hline
\textbf{Sr. no.}                                                   & \textbf{Fig. no.} & \textbf{$R_{\text{solar,SM}}$} & \textbf{$\epsilon_{\text{IR,SM}}$}  \\ \hline
1                                                    &   \ref{fig:NAB_sr4}                                & 0.864 (1.03\%)    & -                                                               \\
\begin{tabular}[c]{@{}l@{}}2\end{tabular}    &     \ref{fig:NAB_sr8}                              & 0.059 (0.01\%)     & 0.710  (8.73\%)                                                  \\
\begin{tabular}[c]{@{}l@{}}3\end{tabular}   &    \ref{fig:AB_sr8}                               & 0.178 (2.73\%)    & 0.391  (6.25\%)                                                  \\
\begin{tabular}[c]{@{}l@{}}4\end{tabular}    &   \ref{fig:AB_sr12}                                & 0.062 (19.2\%)      & 0.935  (5.17\%)                                                  \\ \hline
\end{tabular}
\end{table}

\section{Comparison with experimental data}
\label{sec4:withExptal}
We now apply the semi-analytical technique described in Section \ref{sec:semi_anyl} to predict the optical properties of fabricated coatings reported in literature which have been designed for radiative cooling application. We choose two such disordered coatings where dependent scattering is expected to be dominant so that analytical techniques are not applicable, and the thickness of the coating prohibits the use of exact electromagnetic solvers to predict the optical properties to good accuracy. 

\begin{figure}[ht]
     \begin{subfigure}[b]{0.49\textwidth}
    \centering
    \includegraphics[width=7cm]{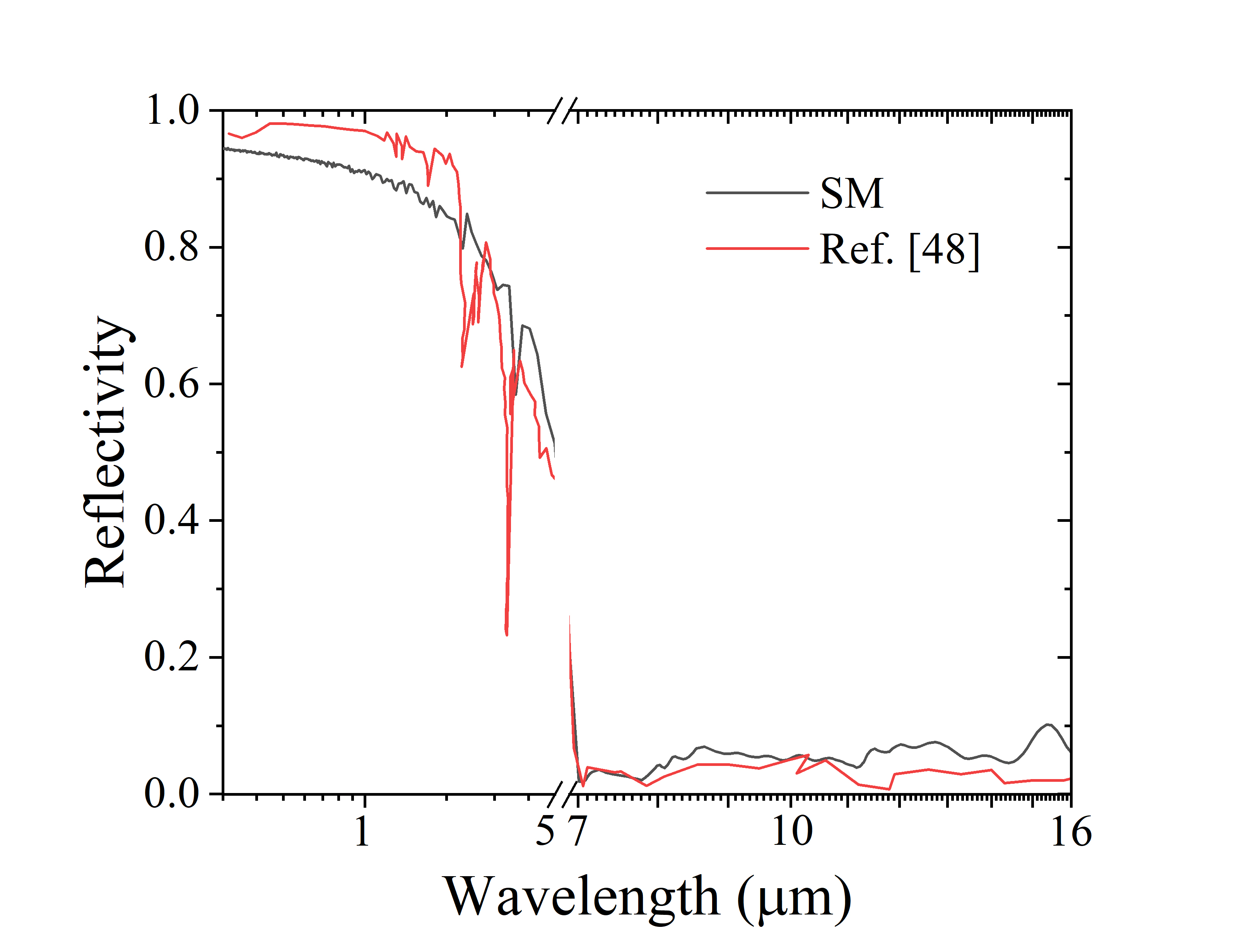}
    \caption{}
    \label{fig:Mandal_SM}
     \end{subfigure}
     \hfill
     \begin{subfigure}[b]{0.49\textwidth}
    \centering
    \includegraphics[width=7cm]{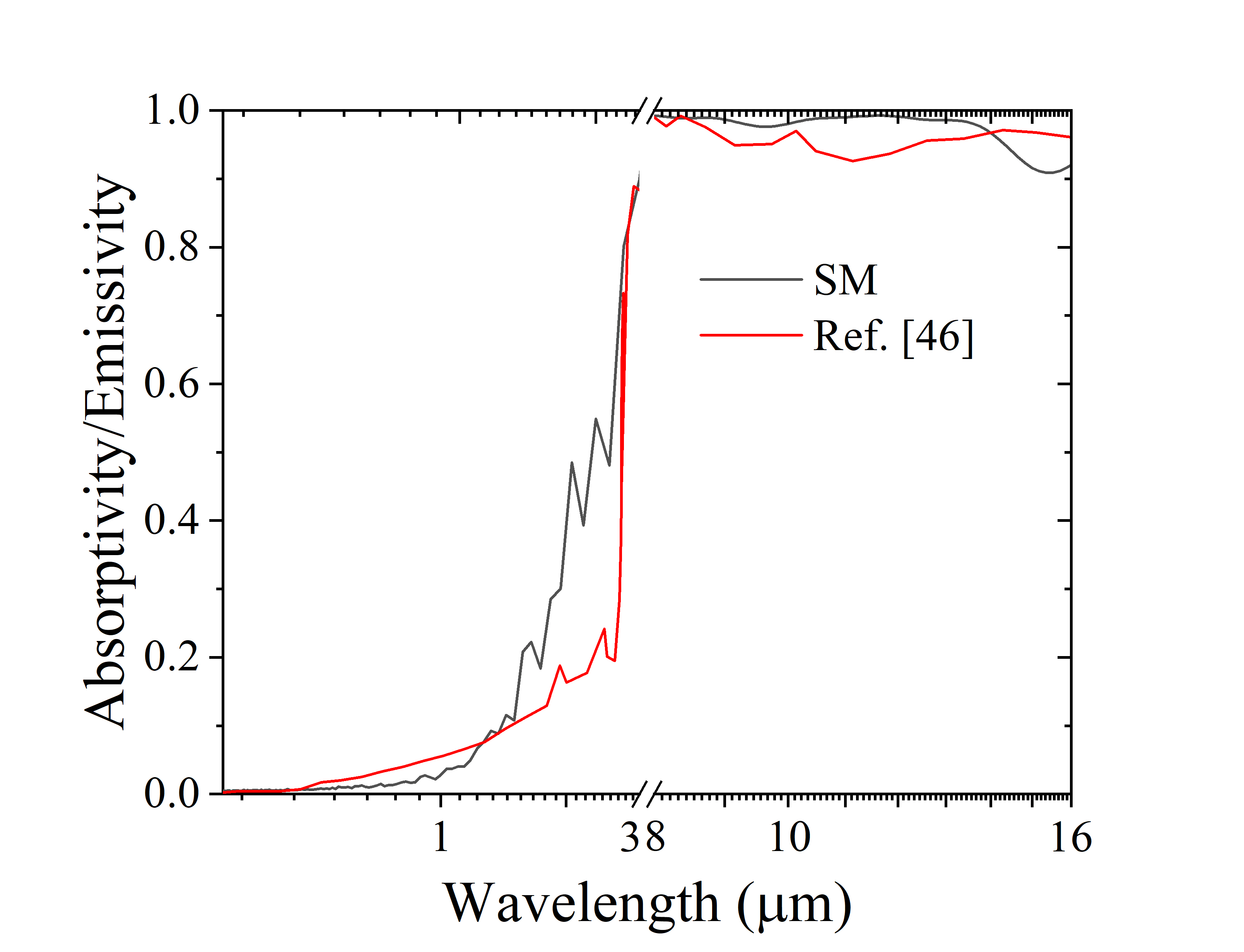}
    \caption{}
    \label{fig:BaSO4_SM}
     \end{subfigure}
    \caption{(a) Reflectivity of hierarchically porous P(VDF-HFP) coating calculated using semi-analytical technique is compared with experimental result given by Mandal et al. \cite{Mandal2018}; (b) Absorptivity/emissivity of BaSO$_4$ film calculated using semi-analytical technique is compared with experimental result given by Li et al. \cite{Li2021baso4}.}
    \label{fig:SM_validation}
\end{figure}

In Ref. \cite{Mandal2018}, a hierarchically porous polymer (P(VdF-HFP)) coating of thickness 300~$\mu$m containing air voids with sizes ranging from 0.05-5~$\mu$m in P(VdF-HFP) matrix has been fabricated, and experimentally characterized to have  solar reflectivity value of 0.96 and emissivity in the 8-13~$\mu$m wavelength range to be 0.97. In order to apply semi-analytical technique to predict the properties of this coating, we set up a simulation in FDTD solver with a smaller coating thickness $t_s = 50~\mu$m (determined using the convergence test explained in Section S4 of supplementary). This thickness is chosen to ensure sufficient number of larger sized air voids  ($r\approx 2.5~\mu$m) in this P(VdF-HFP) matrix. The size distribution of nano-micro air voids used in the simulation is given in supplementary (Fig. S4). Refractive index data of P(VDF-HFP) is extracted from Ref. \cite{Mandal2018}. 
The reflectivity data in the wavelength range $0.3-16~\mu$ m, predicted using the semi-analytical method for $L = 300~\mu$m thickness, is compared with that reported in Ref. \cite{Mandal2018} in Fig. \ref{fig:Mandal_SM}. While an appreciable match is noticed in the predicted values across the spectrum, small deviation observed in the reflectivity values can be attributed to our inability to incorporate exact size distribution of both micro and nano voids as present in the fabricated structure, in ANSYS Lumerical. 

In Ref. \cite{Li2021baso4} an ultrawhite BaSO$_4$ film of thickness 400~$\mu$m has been developed with 60~\% volume fraction of BaSO$_4$ nanoparticles, and has been characterized to have reflectivity of 0.976 in the solar spectrum and emissivity of 0.96 in 8-13~$\mu$m wavelength range. In order to apply the semi-analytical technique to predict the properties of this coating, we set up a simulation in FDTD solver with structure thickness $t_s = 15~\mu$m and BaSO$_4$ spherical particles randomly distributed with volume fraction 60~\%. The particles are taken to be of uniform size distribution with diameters spread over the range $398 \pm 130$ nm to match that reported in Ref. \cite{Li2021baso4}. Matrix is considered to be air for BaSO$_4$ film. Refractive index data of BaSO$_4$ is extracted from Ref. \cite{tong2021atomistic}. 
The emissivity data in the wavelength range $0.3-16~\mu$m, predicted using the semi-analytical method for $L = 400~\mu$m thickness, is compared with that reported in Ref. \cite{Li2021baso4} in Fig. \ref{fig:BaSO4_SM}. While we again observe an appreciable match across the spectrum, some deviation observed particularly around wavelength of  $2~\mu$m is suspected to be due to difference in the refractive index of the fabricated film and that calculated from first-principles in Ref. \cite{tong2021atomistic}.

\section{Conclusion}

In this study we have analyzed the applicability of well-known analytical techniques of KM and FF theories to predict optical properties of a disordered metamaterial coating over a broad spectrum ranging from 300~nm to $15~\mu\text{m}$ wavelength. 
Recent advancements in the use of disordered coatings in  applications such as radiative cooling and solar thermal absorber plates which require tailored optical properties over this wavelength range necessitates such a study. 
%
%
%
Based on deviations observed between the predictions of these analytical techniques and exact FDTD solver in the dependent scattering regime, a two-step semi-analytical technique has been proposed which can be used to predict optical properties of such coatings with good accuracy and minimal computational resources. 
Such a method is expected to be resourceful for designing coatings with specific optical properties where several parameter combinations need to be investigated to arrive at an optimal combination.
Small deviations observed when absorption in host matrix is high  warrants further research to improve this technique. 

\section*{Acknowledgments}
B.R.M. acknowledges support from Prime Minister’s Research Fellowship (PMRF). K.S. acknowledges support from La Fondation Dassault Systèmes and SERB Grant No. SRG/2020/001 511. 
\section*{Disclosures}
The authors declare no conflicts of interest.
\section*{Supplementary information}
See Supplement 1 for supporting content.


\bibliography{Optica-template}

\end{document}